\documentclass[%
 aip,
 amsmath,amssymb,
 reprint,%
]{revtex4-2}

\usepackage{graphicx}% Include figure files
\usepackage{dcolumn}% Align table columns on decimal point
\usepackage{bm}% bold math
\usepackage[utf8]{inputenc}
\usepackage[T1]{fontenc}
\usepackage{mathptmx}
\usepackage{etoolbox}

\usepackage[normalem]{ulem}
\usepackage{color}
\makeatletter
\def\@email#1#2{%
 \endgroup
 \patchcmd{\titleblock@produce}
  {\frontmatter@RRAPformat}
  {\frontmatter@RRAPformat{\produce@RRAP{*#1\href{mailto:#2}{#2}}}\frontmatter@RRAPformat}
  {}{}
}%
\makeatother
\begin{document}

\preprint{AIP/123-QED}

\title[Extreme wave groups on jet currents]{Extreme dynamics of wave groups on jet currents}
% Force line breaks with \\

\author{A.V. Slunyaev}
\email{slunyaev@ipfran.ru.}
\affiliation{%
	National Research University Higher School of Economics, 25 Bolshaya Pechorskaya Street, Nizhny Novgorod 603950, Russia%\\This line break forced% with \\
}%
\affiliation{%
	Institute of Applied Physics, RAS, 46 Ulyanov Street, Nizhny Novgorod 603950, Russia%\\This line break forced% with \\
}%
\affiliation{%
	V.I.Il’ichev Pacific Oceanological Institute, FEB RAS, 43 Baltiyskaya Street, Vladivostok, 690041, Russia%\\This line break forced% with \\
}%

\author{V.I. Shrira}%
 %\email{v.i.shrira@keele.ac.uk.}
\affiliation{School of Mathematics and Computing, Keele University,  ST5 5BG, UK%\\This line break forced with \textbackslash\textbackslash
}%

\date{\today}% It is always \today, today,
             %  but any date may be explicitly specified

\begin{abstract}
Rogue waves are known to be much more common on jet currents. A possible explanation was put forward in Ref.~\onlinecite{ShriraSlunyaev2014PRE}: for the waves trapped on a current robust long-lived envelope solitary waves localized in both horizontal directions become possible, such wave patterns cannot exist in the absence of the current. 
In this work we investigate interactions between envelope solitons of essentially nonlinear trapped waves by means of the direct numerical simulation of the  Euler equations. The solitary waves remain localized in both horizontal directions for hundreds of wave periods. 
We also demonstrate a  high efficiency of the developed analytic nonlinear mode theory for description of the long-lived solitary patterns up to remarkably steep waves. We show robustness of the solitons in the course of interactions, and the possibility of extreme wave generation as a result of solitons' collisions. Their collisions are shown to be nearly elastic. These robust solitary waves obtained from the Euler equations without weak nonlinearity assumptions are viewed as a plausible model of rogue waves on jet currents.

\end{abstract}

\maketitle

%\begin{quotation}
%The ``lead paragraph'' is encapsulated with the \LaTeX\
%\verb+quotation+ environment and is formatted as a single paragraph before the first section heading.
%(The \verb+quotation+ environment reverts to its usual meaning after the first sectioning command.)
%Note that numbered references are allowed in the lead paragraph.
%
%The lead paragraph will only be found in an article being prepared for the journal %\textit{Chaos}.
%\end{quotation}

\section{\label{sec:Introduction} Introduction}

Rogue waves in the ocean are known to be much more common on currents, the Agulhas current near the south-eastern coast of Africa being a notorious example \cite{Mallory1974,Kharifetal2009}. To explain this prevalence at first  a number of  linear mechanisms of wave amplification by currents were suggested  (see Refs.~\onlinecite{Peregrine1976,Smith1976,WhiteFornberg1998}). Later on, the role of nonlinearity got the most of attention.  Simulations of nonlinear  wave evolution on the Agulhas current based on the kinetic equation with account of wave refraction on the current were carried out in Ref.~\onlinecite{Lavrenov1998}.

However,  the main focus was on studies of modification by the current   of the Benjamin--Feir (modulational) instability,  since for narrowband wavefields of a given significant wave height the  modulational instability increases the probability of high waves. To this end various versions of nonlinear Schr\"odinger equations (NLSE) were derived
and analyzed  under assumptions of slow current, weak nonlinearity, and narrow-banded spectrum (e.g.  Ref.~\onlinecite{HjelmervikTrulsen2009}, and
references therein). The Benjamin--Feir instability was found to be strengthened for waves on adverse intensifying currents, which was also confirmed in numerical simulations \cite{JanssenHerbers2009,Onoratoetal2011,Ruban2012}. 

All the above works were concerned with free water waves propagating on horizontally inhomogeneous currents, however, besides free waves ocean currents  also support the existence of trapped waves which we expect to play a key role in explaining the observed prevalence of rogue waves on currents.  In the real ocean the characteristic nonlinear spatial  scales  are such that the traditional ray theory cannot faithfully describe the evolution of nonlinear trapped waves (see reviews in Refs.~\onlinecite{HjelmervikTrulsen2009,ShriraSlunyaev2014JFM}).  To address  this gap in the mathematical theory,  a novel approach enabling one to describe linear trapped waves asymptotically was put forward in  Ref.~\onlinecite{ShriraSlunyaev2014JFM}. Utilizing the found asymptotic solutions for trapped modes, a
  weakly nonlinear mode approach was developed in Ref.~\onlinecite{ShriraSlunyaev2014PRE}. 
Nonlinear dynamics of  waves trapped by an opposing jet current was  studied both analytically and
numerically. It was found that for wave fields narrowband in frequency, but not necessarily with narrow angular distributions, the wave dynamics  for a single mode is described to leading order by the 1D  modified nonlinear  Schr\"odinger equations equation of self-focusing
type.

The NLSE  solutions, such as envelope solitons and breathers,  are commonly considered to be prototypes of
rogue waves.  Ongoing intense research effort is dedicated to the investigation of  complex interactions between envelope solitons and breathers within the framework of the integrable NLSE and its non-integrable generalizations, including the strongly nonlinear frameworks, see Refs.~\onlinecite{Akhmedievetal2016,Kachulinetal2020,Suretetal2020,Slunyaev2021,Slunyaevetal2023} and references therein, among many others. 

Crucially, in contrast to the case of free waves in the absence of currents, both solitons and breather prototype  solutions of trapped waves are robust with respect to transverse perturbations. The linear trapped mode solutions are not exact solutions, which results in  a slow leakage of energy from the mode under consideration. Direct numerical
simulations of the full potential Euler equations showed that generalizations of these solutions satisfactorily describe long-lived structures.    The fact that these structures are long-lived suggests an increased likelihood  of encountering their real world counterparts. One of the important issues left open in Ref.~\onlinecite{ShriraSlunyaev2014PRE} is the robustness of the found structures with respect to collisions with other essentially nonlinear structures.

In this work we investigate interactions between envelope solitons of essentially nonlinear trapped waves by means of the direct numerical simulation of the  Euler equations. The solitary waves remain localized in both horizontal directions. We also demonstrate the high efficiency of the developed analytic nonlinear mode theory for description of the long-lived solitary patterns, even for remarkably steep waves. We show the robustness of the solitons with respect to interactions, and the possibility of extreme wave generation as a result of collisions of envelope solitons. The robust solitary waves obtained from the Euler equations without weak nonlinearity assumptions are viewed as a plausible model of rogue waves on jet currents.

\section{\label{sec:PrimitiveEquations} Governing equations and numerical approach}

We study wave motions on the free surface of ideal incompressible fluid of unit density under the action of gravity. Waves are propagating  along the Ox axis on a given steady current $\textbf{U}$,
which depends only on the transverse coordinate $y$, $\textbf{U}=(U(y),0,0)$.  The assumption of potential wave motions applies, so that the full Eulerian velocity of the fluid reads 
$\textbf{v}(x,y,z,t) = \textbf{U}+\nabla \varphi(x,y,z,t)$, where $\varphi$ is the velocity potential related to the wave perturbations. The water surface is specified by the condition 
$z=\eta(x,y,t)$ with the rest level $z=0$, where the axis Oz is directed upward; the water depth is infinite. The set of hydrodynamic equations may be formulated \cite{Zakharov1968}
 in the form of two conditions on the free surface,
\begin{subequations}\label{PrimitiveEquations}
	\begin{align}
		\eta_t + (\nabla \Phi + \textbf{U}) \cdot \nabla \eta = \varphi_z \left( 1 + {\nabla \eta}^2 \right) , \quad z=\eta,\\
		\Phi_t + g \eta + \frac{1}{2}(\nabla \Phi + \textbf{U})^2 + P = \frac{1}{2} {\varphi_z}^2 \left( 1 + {\nabla \eta}^2 \right), \quad z= \eta ,
	\end{align}
\end{subequations}
and the Laplace equation in the water column,
\begin{eqnarray} \label{Laplace}
	\nabla^2\varphi + \varphi_{zz} = 0, \quad z \le \eta \,.
\end{eqnarray}%
The equations are complemented by the condition of motion decaying at infinity
\begin{align} \label{BottomBC}
\varphi \to 0 ,\quad z \to - \infty. 	
\end{align}
In Eqs.~(\ref{PrimitiveEquations}) and (\ref{Laplace}), the gradient operator acts in the horizontal plane only, $\nabla \equiv (\partial / \partial x, \partial / \partial y)$, $g = 9.81$~ms$^{-2}$ is the gravity acceleration, $P(x,y,t)$ is the pressure applied to the water surface.
Note that due the specific form of the current $\textbf{U}$, the incompressibility condition reduces to the standard Laplace equation (\ref{Laplace}).

Under the stationary rest condition $\eta=0$, $\varphi=0$, the primitive equations (\ref{PrimitiveEquations}) yield the relation on pressure
\begin{eqnarray} \label{RestWaterBalance}
	\overline{P} = - \frac{1}{2} U^2,
\end{eqnarray}%
which specifies spatially inhomogeneous pressure needed to ensure the unperturbed water surface upon background jet current at  $z=0$. Then, the total surface pressure may be represented in the form
$P(x,y,t) = \overline{P}(y) + P_{atm}(x,y,t)$, where $P_{atm}$ is the atmosphere pressure. 
Assuming the atmosphere pressure to be constant, the original equations on the free surface (\ref{PrimitiveEquations}) may be re-written in the equivalent form
\begin{subequations}\label{PrimitiveEquations2}
	\begin{align}
		\eta_t + (\nabla \Phi + \textbf{U}) \cdot \nabla \eta = \varphi_z \left( 1 + {\nabla \eta}^2 \right) ,\quad z=\eta,\\
		\Phi_t + g \eta + U \Phi_x + \frac{1}{2}(\nabla \Phi )^2 = \frac{1}{2} {\varphi_z}^2 \left( 1 + {\nabla \eta}^2 \right), \quad z=\eta \,.
	\end{align}
\end{subequations}
The set of of equations (\ref{Laplace}), (\ref{BottomBC}) and (\ref{PrimitiveEquations2}) specifies  the Cauchy problem in a closed form.
In this work, this system is solved numerically using an adapted version of the High Order Spectral Method \cite{Westetal1987}.
%, which is briefly described in the Appendix.

The general solution for the velocity potential $\varphi(x,y,z,t)$ is represented in terms of exact solutions of the Laplace equation (\ref{Laplace}) with defined values at the water rest level $\varphi(x,y,z=0,t)$ and vanishing at great depths (\ref{BottomBC}). At the horizontal plane $z=0$ this representation coincides with the double Fourier transformation with respect to $x$ and $y$. The surface velocity potential $\Phi(x,y,t)$ and the vertical velocity on the water surface $\varphi_z(x,y,z=\eta,t)$ are linked with the values $\varphi(x,y,z=0,t)$ using the Taylor expansions of the order $M$ near the horizon $z=0$.
Thus, the method is fully accounting  dispersive effects, but is somewhat limited in nonlinearity. It describes accurately the nonlinear interactions between up to $M-1$ waves \cite{Onoratoetal2007}. In this work the nonlinearity parameter was set $M=5$, which allows to simulate up to 6-wave interactions. Thus, the chosen setting corresponds to a strongly nonlinear framework.

The governing hydrodynamic equations possess a set of conservation integrals:  mass $\cal{M}$, flux $\cal{F}$, momentum $\textbf{\cal{P}}$ and transformation of the mechanical energy $\cal{W}$, which for the equations with the current $\textbf{U}=(U(y),0,0)$ have the following forms:
\begin{eqnarray}
	{\cal{M}} = \int{\eta dx dy} = Const, \label{Mass} \\
	{\cal{F}} = \int{\eta_t dx dy} = Const, \label{Flux} \\
	\textbf{{\cal{P}}} = \int{\eta \left( \textbf{U} + \nabla \Phi \right) dx dy} = Const, \label{Moment} \\
	{\cal{E}}={\cal{W}} +  \int_0^t{{\cal{A}}(\tau) d\tau} = Const. \label{EnergyLaw}	
\end{eqnarray}%
The potential and kinetic constituents of the energy, ${\cal{W}}_p(t)$ and ${\cal{W}}_k(t)$, ${\cal{W}}(t) = {\cal{W}}_p+{\cal{W}}_k$, and the action ${\cal{A}}(t)$ due to the current read:
\begin{eqnarray}
	{\cal{W}}_p = \frac{1}{2} \int{g \eta^2 dx dy}, \label{Ep} \\
	{\cal{W}}_k = \frac{1}{2} \int{ \left( \Phi \eta_t + \eta U^2 - \Phi U \eta_x \right) dx dy}, \label{Ek} \\
	{\cal{A}} = \int{\overline{P} \eta_t dx dy} \label{Action}.	
\end{eqnarray}%

The numerical integration in time is performed using the 4-th order Rounge-Kutta method with the constant step $0.125$~s. The accuracy of simulations is controlled by checking the energy quantity  ${\cal{E}}(t)$, which should be preserved in a perfect simulation.

In the computer experiments discussed in Sec.~\ref{sec:2SolInteractions} the size of the simulation domain is $L_x \times L_y = 6400 \pi \times 400\pi$ square meters with the spatial resolution $2^{13}\times 2^8$ points; the relative error $|{\cal{E}}(t)-{\cal{E}}(0)|/{\cal{E}}(0)$ in all simulations is less than $2\cdot 10^{-4}$.
The number of grid points in the corresponding Fourier domain is doubled along each coordinate in order to partly solve the aliasing problem. In the simulation discussed in Sec.~\ref{sec:DegenerateSolitons} the length of the domain $L_x$ is twice shorter with the same spatial resolution; the relative error is within $3.2\cdot 10^{-4}$.

%In all presented simulations the relative error $|{\cal{E}}(t)-{\cal{E}}(0)|/{\cal{E}}(0)$ was less than $5\cdot 10^{-4}$. The spatial resolution is $2^{13}\times 2^8$ points along the longitudinal and transverse coordinate, respectively, in the simulations discussed in Sec.~\ref{sec:2SolInteractions}, and $2^{12}\times 2^8$ points in Sec.~\ref{sec:DegenerateSolitons}.

\section{\label{sec:CurNLSE} Nonlinear framework for trapped modes}

The modal theory for waves traveling on the jet current may be developed using uniformity of the rest conditions with respect to $x$ and $t$, see Ref.~\onlinecite{ShriraSlunyaev2014JFM}. Then, within the linear approximation the full solution represents a superposition of traveling waves with longitudinal wavenumbers and angular frequencies $(k,\omega)$ propagating collinear to the current with some mode structure in the $(y,z)$ plane.

The lateral modes are described by a two-dimensional boundary value problem (BVP). Under the assumption that the current is broad compared to the longitudinal wave length, so that the 
wave is described locally by the solution for a uniform current, the BVP may be reduced to the following  one-dimensional nonlinear boundary value problem
\begin{align} \label{BVP1D}
	\frac{d^2 Y}{d y^2} + \frac{k^2}{\omega_g^4} \left(\Omega^4 - \omega_g^4 \right) Y =  0, \\
	\omega_g = \sqrt{k g}, \qquad \Omega(y) = \omega - kU , \nonumber
\end{align}
where $\omega_g$ denotes the frequency of linear gravity waves in still water, and  $Y(y)$ is the transverse mode.  The decaying or non-decaying boundary conditions on $Y(y)$ as $x\to\pm \infty$ specify trapped and passing-through modes, respectively.

 Upon multiplication of (\ref{BVP1D}) by $Y(y)$, consequent integration with respect to $y$, $y\in [a,b]$, and  integration by parts, we obtain that $YY^\prime |_{a}^{b} - \int_a^b{{Y^\prime}^2dy} + \frac{k^2}{\omega_g^2} \int_a^b{(\Omega^4-\omega_g^4)Y^2dy} = 0$.
If $\frac{d}{dy}Y^2$ decays to zero when $y=a$ and $y=b$ (for example, in infinite line, when $a\to-\infty$ and $b\to +\infty$) or if the problem is periodic, $\frac{d}{dy}Y(a)^2=\frac{d}{dy}Y(b)^2$, then the BVP (\ref{BVP1D}) has no solution if $\Omega^4 < \omega_g^4$ for all $y \in [a,b]$.
Far outside the current, when one may put $U=0$, the modes $Y(y)$ decay when $\omega < \omega_g$.
Therefore, trapped modes may occur only in the frequency interval $\omega_C < \omega < \omega_g$, $\omega_C= \omega_g+\min{(kU)}  <\omega_g$, when the current is opposite to the direction of wave propagation. The frequency $\omega_C$ corresponds to the frequency with the Doppler shift produced by the adverse uniform current of the speed $\max{|U|}$. When the current is strictly opposite, $kU(y)<0$, the term $\Omega(y)$ is never zero.

For a given $k$ the eigenproblem (\ref{BVP1D}) is nonlinear in terms of the frequency $\omega$. For a given $k$, it may have more than one eigenvalue and eigenfunction, which we will label with the subscripts: $\omega \in \{\omega_n\}$, $Y \in \{ Y_n\}$.
In case of infinite domain, it is straightforward to show \cite{ShriraSlunyaev2014JFM},  that within the BVP (\ref{BVP1D}) each eigenfrequency corresponds to unique eigenfunction for trapped modes, and to two linearly independent eigenfunctions for passing modes.

The  BVP (\ref{BVP1D}) may be further simplified under the assumption of a weak current, $|kU/\omega| \ll 1$. Then it reduces to the linear Sturm--Liouville type problem on the eigenfrequency $\omega$
\begin{align}  \label{BVP-SL}
	\frac{d^2 Y}{d y^2}  + 4 \frac{k^2}{\omega_g} \left[ \omega - \left( \omega_g + kU \right) \right] Y =0.
\end{align}
Similar to the BVP (\ref{BVP1D}), trapped modes may appear only in the frequency interval $\omega_C < \omega < \omega_g$.

For the Sturm--Liouville BVP (\ref{BVP-SL}) on the whole axis  $-\infty < y < \infty$ the combination of trapped modes (which belong to the discrete spectrum) and non-localized modes 
(which belong to the continuous spectrum) form a complete basis of functions $Y_n(y)$ which are orthogonal in the sense
\begin{align}  \label{Orthogonality}
	N_{nm} = \frac{\int{Y_n(y) Y_m(y) dy}}{\int{Y_n^2(y)dy}} = \delta_{n,m},
\end{align}
where $\delta_{n.m}$ is the Kronecker delta, and thus provide a convenient base for studying waves on jet currents.

For the Sturm--Liouville BVP in periodic domain, $U(y+L_y)=U(y)$, the set of eigenfrequencies ${\omega_n \in \Re}$ is discrete and infinite; the system of eigenfunctions is full and may be chosen orthogonal in the form (\ref{Orthogonality}), where the integration is performed over the interval of periodicity $L_y$.

For a given longitudinal wavenumber $k$, the general solution for the water surface displacement is given by the superposition of the modes:
\begin{equation} \label{SurfaceReconstruction}
	\eta(x,y,t) =  \sum_n{ \psi_n Y_n(y) \exp{\left( i \omega_n t - i k x \right)}} + c.c.,
\end{equation}
where $\psi_n$ are the mode amplitudes. Hereafter, we will use the mode normalization assuming its maximum displacement is equal to one,
\begin{equation} \label{ModeCalibration}
	\max{|Y_n(y)|}=1 \quad \text{for all} \,\, n,
\end{equation}
so that the maximum mode amplitude $|\psi_n|$ coincides with the corresponding maximum surface displacement $|\eta|$.
Note that consideration of a different wavenumber $\tilde{k}$ will lead to a new BVP (\ref{BVP-SL}) with new eigenfrequencies $\tilde{\omega}_m$ and corresponding set of orthogonal functions $\tilde{Y}_m(y)$, which should be added to the representation (\ref{SurfaceReconstruction}).
Since the modes may evolve, the amplitudes are slow functions of time and longitudinal coordinate, $\psi_n=\psi_n(x,t)$.

This way, a rigorous asymptotic theory for weakly nonlinear slow modulations may be developed \cite{ShriraSlunyaev2014JFM,ShriraSlunyaev2014PRE}, which yields nonlinear evolution 
equations on the amplitudes of certain modes of the BVP for a given longitudinal wavenumber. The third-order in nonlinearity dispersive theory for a single mode trapped by a weak current was
formulated in Ref.~\onlinecite{ShriraSlunyaev2014PRE}, which has the form of the nonlinear Schr\"odinger equation. For the complex amplitude $\psi_n(x,t)$ of the $n$-th mode with the 
longitudinal carrier wavenumber $k$ the nonlinear evolution equation is
\begin{align} \label{NLS}
	 \frac{i}{\omega_n} \left( \frac{\partial \psi_n}{\partial t} +  V_n \frac{\partial \psi_n}{\partial x} \right) +
	  \frac{1}{8k^2} \frac{\partial^2 \psi_n}{\partial x^2} + \frac{ k^2}{2}  I_n^2  \psi_n |\psi_n|^2 +  \\
	  +k^2 \sum_j{\mu_{j} I_{nj}^2 \psi_n |\psi_j|^2} + \sum_{p,q,r}{\nu_{npqr} J_{npqr} \psi_p^* \psi_q \psi_r}=0, \nonumber \\
	%    V = \frac{\omega_n}{2k} + U(y),
	V_n= \frac{1}{  \int_{-\infty}^{\infty}{Y_n^2 dy}}
	\int_{-\infty}^{\infty}{\left( \frac{k g^2}{2 \Omega_n^3} + U \right) Y_n^2 dy} \,, \nonumber \\
		\mu_{j} = 
\left\{
\begin{array}{ll}
	\sqrt{\frac{k}{k_j}}, & \hbox{if $k_j > k$,} \\
	\sqrt{\frac{k_j}{k}}, & \hbox{if $k_j < k$,}
\end{array} 
\right.  \nonumber
\\
%	\mu_{j} = \left( \frac{k_j}{k} \right)^{{\text{sgn}(k-k_j)}/{2}} , \nonumber \\
	I_n^2 = \frac{  \int_{-\infty}^{\infty}{  Y_n^4 dy}}{  \int_{-\infty}^{\infty}{Y_n^2 dy}}, \quad
	I_{nj}^2 = \frac{  \int_{-\infty}^{\infty}{  Y_j^2 Y_p^2 dy}}{  \int_{-\infty}^{\infty}{Y_n^2 dy}} , \nonumber \\
	J_{npqr} = \frac{  \int_{-\infty}^{\infty}{  Y_n Y_p Y_q Y_r dy}}{  \int_{-\infty}^{\infty}{Y_n^2 dy}}.
		\nonumber
\end{align}
Here $I_n^2\le 1$ is the mode overlap integral which determines the coefficient of the $n$-th mode self-interaction. 
The first term in the second line of the evolution equation (\ref{NLS}) is responsible for trivial resonant interactions. For every index $j$, $\psi_j(x,t)$ are complex amplitudes of waves with the wavenumber $k_j > 0$ and lateral modes $Y_j$ which are solutions of the BVP for $k_j$. The trivial interactions do not lead to the energy exchange between modes, $\frac{d}{dt} \int_{-\infty}^{\infty}{|\psi_n|^2 dx} = 0$.
The nonlinear coefficient for trivial interactions coincides (except for the integral parameter $I_{nj}$) with the one obtained in Ref.~\onlinecite{Lavrova1983} for the problem of two interacting wave systems in still water.

The last term in the equation (\ref{NLS}) describes non-trivial resonant interactions with wavenumbers $k_p>0$, $k_q>0$, $k_r>0$ and corresponding frequencies $\omega_p$, $\omega_q$, $\omega_r$, so that $k+k_p=k_q+k_r$ and $\omega_n+\omega_p=\omega_q+\omega_r$.
The functions $\psi_p(x,t)$, $\psi_q(x,t)$, $\psi_r(x,t)$ are amplitudes of resonant quartets which belong to modes $Y_p$, $Y_q$ and $Y_r$ respectively, -- solutions of corresponding BVPs.
The coefficient of non-trivial interactions $\nu_{npqr}$ has complicated form and is not given in this paper. The overlap integrals $J_{npgr}$ are assumed to be generally small.

The nonlinear and dispersion coefficients in (\ref{NLS}) are written in the limit of a weak current, under which the expression for velocity may be also simplified as $V_n=\frac{kg^2}{2 \omega_n^3}+\frac{5}{2} \overline{U}$, where $\overline{U}=\int{UY_n^2dy}/\int{Y_n^2dy}$. The two contributions to $V_n$ depend on the mode number and are of different signs. Hence the dependence of $V_n$ on the mode number may be nontrivial.

%Note that differen longitudinal wavenumbers determine different BVPs and different sets of orthogonal functions in (\ref{BVP-SL}).

If the interactions with other modes are disregarded, which can be justified for a wide range of situations, the evolution equation (\ref{NLS}) differs from the classical NLSE in still water by the account for the effective Doppler frequency shift and a reduced
nonlinear coefficient  due to factor $I^2_n < 1$. It is known to be integrable, they key role in the evolution is played by envelope solitons.

The envelope soliton solution may be written as
\begin{align} \label{OneSoliton}
\psi_{\text{es}}(x,t) =A \frac{\exp{\left( i \frac{k^2 A^2}{4} I_n^2 \omega_n t \right)}}{\cosh{\left[ \sqrt{2} k^2 A I_n \left( x-V_n t\right) \right]}} = \\
=\frac{{a}}{I_n} \frac{\exp{\left( i \frac{k^2 {a}^2}{4}  \omega_n t \right)}}{\cosh{\left[ \sqrt{2} k^2 {a} \left( x-V_n t\right) \right]}} , \nonumber
\end{align}
where $A$ and $a = I_n A$ are amplitude parameters. One may see, that the effect of the jet current on the soliton longitudinal structure may be reduced to the presence of a calibration constant for the soliton amplitude ${a}$ in the problem with no current. Due to the normalization of modes by their maximum values (\ref{ModeCalibration}), the parameter $A$ corresponds to the observable amplitude of the water displacement. Since $A= a /I_n$ and $I_n<1$, the amplitudes of solutions of the NLSE upon jet current will be increased compared to the solution of the NLSE without current by the factor $I_n^{-1} > 1$. For the example of the current which we consider in this work (see the next section), we have $I_1^{-1} \approx 1.19$, $I_2^{-1} \approx 1.18$, $I_3^{-1} \approx 1.21$. Respectively, the width of the envelope soliton with a given wavenumber and a given maximum amplitude is larger in the case with opposite current compared to still water.

\section{\label{sec:CurrentInDNLS} The jet current condition in the numerical simulations}

\begin{figure}[h]
	\includegraphics[width=7cm]{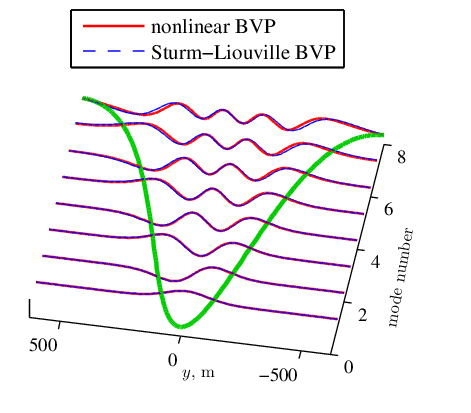}(a) \\ % Here is how to import EPS art
	\includegraphics[width=7cm]{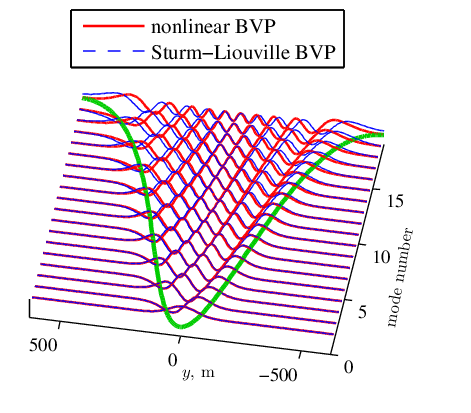}(b)% Here is how to import EPS art
	\caption{\label{fig:TrappedModes} Eigenmodes for the longitudinal wavenumbers $k=0.05$~rad/m (a) and  $k=0.1$~rad/m (b). The solid red and dashed blue curves correspond to the modes $Y_n$ obtained solving the nonlinear BVP (\ref{BVP1D}) and the Sturm--Liouville BVP (\ref{BVP-SL}), respectively. The green curves show the jet current profile. }
\end{figure}

In order to apply the pseudo-spectral numerical scheme, periodic boundary conditions are imposed in both, $x$ and $y$ coordinates. For certainty and convenience we choose the jet current 
to be of a cnoidal shape,
\begin{align} \label{CurrentShape}
U = U_0 \text{cn}^2{(2K \frac{y}{L_y},s^2)},
\end{align}
where $K(s^2)$ is the complete elliptic integral of the first kind with the parameter $s$. We choose the parameters of our model current to be: $U_{0}=-2$~m/s, $L_y=400 \pi \approx 1200$~m  and $s=0.9$.  Thus, the current varies from $U=0$ at $y=\pm L_y/2$ to $U=U_{0}$ at $y=0$ as shown by green curves in Fig.~\ref{fig:TrappedModes}.

In this work we operate with dimensional variables; the parameters of the jet current and of the waves chosen to be realistic.

The BVPs (\ref{BVP1D}) and (\ref{BVP-SL}) are solved numerically using the spectral method with the help of the convolution theorem. Then the problem is reduced to a homogeneous system of $N$ linear algebraic equations on $N$ Fourier amplitudes of the mode $Y_n(y)$, which may have solution only when its determinant is zero. Thus, the eigenvalue problem is solved by minimizing the determinant using the shooting method for frequencies in the interval of trapped waves, $\omega_g-|k{U_0}| < \omega < \omega_g$. When the eigenfrequencies $\{ \omega_n \}$ are found, the corresponding eigenmodes $\{ Y_n \}$ are obtained by solving the inhomogeneous system of $N-1$ linear algebraic equations on $N-1$ Fourier amplitudes of the function $Y_n$. The solutions of the BVPs  obtained in this manner are accurate to the computer precision.

\begin{figure}[h]
	\includegraphics[width=7cm]{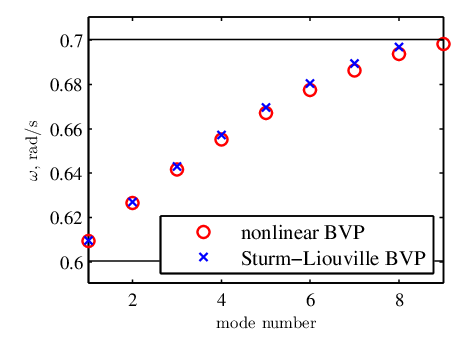}(a) \\ %NLSsoliton3DCur7@Umax=2@kx=0.05@EigenFrequenciesFullvsSL.eps}(a) \\ % Here is how to import EPS art
	\includegraphics[width=7cm]{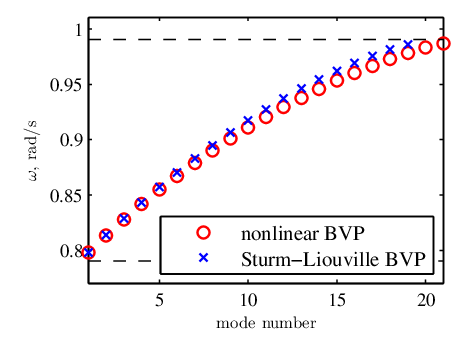}(b) %NLSsoliton3DCur7@Umax=2@kx=0.1@EigenFrequenciesFullvsSL.eps}(b)% Here is how to import EPS art
	\caption{\label{fig:Frequencies} Eigenfrequencies for the same conditions as per Fig.~\ref{fig:TrappedModes}~a,b. The upper and lower horizontal dashed lines correspond to the frequencies $\omega_C$ and $\omega_g$, respectively.}
\end{figure}

Solutions of the BVPs for the longitudinal wavenumbers $k=0.05$~rad/m and $k=0.1$~rad/m are displayed in Fig.~\ref{fig:TrappedModes} and Fig.~\ref{fig:Frequencies}.
The mode shapes obtained within the nonlinear BVP (\ref{BVP1D}) and the linear Sturm--Liouville problem (\ref{BVP-SL}) are close, especially for the modes with low numbers. Hereafter the modes are sorted in the order of ascending eigenfrequencies. Frequencies of low modes obtained for the problems (\ref{BVP1D}) and (\ref{BVP-SL}) are very close, see Fig.~\ref{fig:Frequencies}. At the same time, the total number of modes with the frequencies in the interval of trapped waves  $[\omega_C,\omega_g]$ may be different. The closeness of the solutions of the two BVPs may be explained by the smallness of the ratio of the current velocity to the wave phase speed, $|kU/\omega| \lesssim 0.25$.
%, which is larger in value for the case of shorter waves $k=0.1$~rad/m.

  It is illuminating to use the quantum mechanics analogy for the BVP (\ref{BVP-SL}), where the opposing current plays the role of a potential well, while the energy of the quantum system is proportional to $(\omega-\omega_g)<0$. Then the low-number modes correspond to the least energetic states, which are determined by configuration of the well bottom (i.e., by the shape of the jet current tip).
If the simulated domain is extended in the transverse direction, the chosen shape (\ref{CurrentShape}) may be approximated by a sech function, and the modes with low numbers will be close to the modes of trapped waves of the problem on the full axis. At the same time, in the case of a finite width of the simulated domain, modes with high numbers are obviously affected by the periodic condition along the transverse coordinate. In what follows we focus upon the dynamics of low-mode waves,  hence, we may say with some justification that we study trapped waves.

\begin{figure}[h]
	\includegraphics[width=8cm]{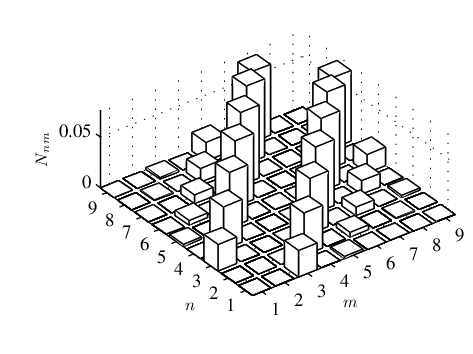}(a) \\ %JetCurrentBVP@Umax=2@kx=0.05@Method=1DFull@CrossOrthogonality.eps}(a) \\ % Here is how to import EPS art
	\includegraphics[width=8cm]{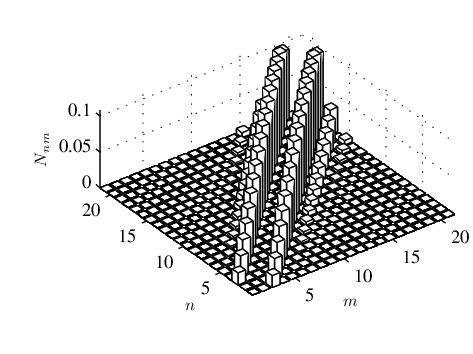}(b) %JetCurrentBVP@Umax=2@kx=0.1@Method=1DFull@CrossOrthogonality.eps}(b)% Here is how to import EPS art
	\caption{\label{fig:Inp} Overlap integrals $N_{nm}$ for trapped modes with indices $n$ and $m \ne n$ for the longitudinal wavenumbers $k_0=0.05$~rad/m (a) and  $k_0=0.1$~rad/m (b); the values $N_{nn}=1$ are not shown.}
\end{figure}

\begin{figure}[h]
	\includegraphics[width=7cm]{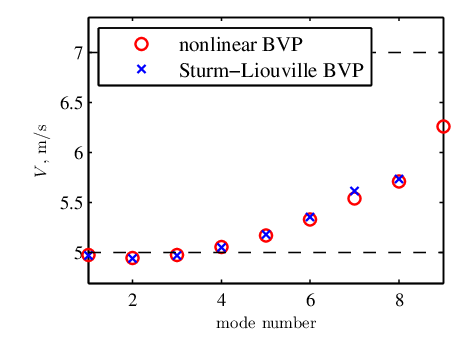} %NLSsoliton3DCur7@Umax=2@kx=0.05@ModeVelocitiesFullvsSL.eps}
	\caption{\label{fig:ModeVelocities} Mode velocities for the longitudinal wavenumber $k_0=0.05$~rad/m. The upper and lower horizontal dashed lines correspond to the reference velocities  $\frac{\omega_g}{2k}$ and $\frac{\omega_g}{2k}-|U_0|$, respectively.}
\end{figure}

The modes within the framework of  the nonlinear BVP (\ref{BVP1D}) are not orthogonal, but the overlap integrals $N_{nm}$ for different modes are small,  as shown in Fig.~\ref{fig:Inp}. Mode velocities $V_n$, calculated according to the definition in ($\ref{NLS}$) demonstrate non-monotonic dependence on the mode number for a given $k$, see Fig.~\ref{fig:ModeVelocities}. As a result, the celerities of low modes are particularly close.

\section{\label{sec:2SolInteractions} Collisions of trapped envelope solitons}

The initial conditions for the simulations of pairs of solitons are taken in the form of a linear superposition of two independent solitons according to (\ref{OneSoliton}) and (\ref{SurfaceReconstruction}), and the linear theory for the velocity potential. Accordingly, the surface displacement and the surface velocity potential are set at $t=0$ as follows:
\begin{subequations}\label{InitialCondition}
\begin{align} \label{InitialConditionEta}
	\eta(x,y,0) = \frac{1}{2} \sum_{j=1}^2{\psi_{0,j}(x) Y_j(y) \exp{\left( i \omega_j t - i k_j x \right)}} + c.c. , \\
	\Phi(x,y,0) = \frac{1}{2}  \sum_{j=1}^2{ \frac{ig}{\omega_g} \psi_{0,j}(x) Y_j(y) \exp{\left( i \omega_j t - i k_j x \right)}} + c.c., \label{InitialConditionPhi}
\end{align}
\end{subequations}
where $\psi_{0,j}= \psi_{\text{es},j}(x,t=0)$, $j=1,2$. Several modifications of the initial condition for the surface velocity potential were tested through numerical simulations. According to the analytic theory, $\Omega_n(y)$ should be placed in the denominator in (\ref{InitialConditionPhi}) instead of $\omega_g$. However, when it is replaced by $\omega_g$, the generated solitary structures are found exhibiting less radiation at the initial stage of simulations.

The solitons are characterised by the carrier longitudinal wavenumbers $k_j$; frequencies $\omega_j$ and modes $Y_j(y)$ which correspond to the mode sequence numbers $n_j \ge 1$; and amplitudes $A_j$, see the parameters of experiments in Table~\ref{tab:SimulationParameters}. A simulation of an interaction between two solitons of the mode $n$ (for longer waves) and $m$ (for shorter waves) is referred to as ''S$n$-$m$``.
In order to facilitate the interpretation of the simulation results, each run of interacting solitons (i.e., runs S1-1, S2-1, S3-1) was accompanied by additional numerical simulations, where single solitons were modeled (runs S1-0, S0-1, S2-0, S3-0).
The solitons differ in the carrier wavelengths and the amplitudes by a factor of two, so that the maximum steepness of the solitons are similar, $k_1A_1 = k_2A_2 = 0.2$.

\begin{table}
	\caption{\label{tab:SimulationParameters}Parameters of numerical simulations. The wavenumbers $k_j$ are in rad/m; the surface displacement amplitudes $A_j$ are in meters, the mode periods $T_{n_j}$ are is seconds, and mode speeds $V_j$ are in m/s, $j=1,2$. Solutions of the nonlinear problem (\ref{BVP1D}) are used.}
	\begin{ruledtabular}
		\begin{tabular}{ccccccccccc}
			No & $k_1$ & $A_1$ & $n_1$ & $T_{n_1}$ & $V_1$ &  $k_2$ & $A_2$ & $n_2$ &$T_{n_2}$ & $V_1$\\
			\hline
			S1-1&0.05 & 4 &1 & 10.3 &4.98 &0.1 & 2 &1 &7.9 & 2.94 \\
			S1-0 &0.05 & 4 &1 &10.3  &4.98 &-- & -- &-- &-- & --\\
			S0-1 &-- &--   &-- &--  &-- &0.1 & 2 &1 &7.9  & 2.94 \\
			S2-1 &0.05 & 4 &2 &10.0  &4.95 &0.1 & 2 &1 &7.9  &2.94 \\
			S2-0 &0.05 & 4 &2 &10.0  &4.95 &-- & -- &-- &--  &-- \\
			S3-1 &0.05 & 4 &3 &9.8  &4.98 &0.1 & 2 &1 &7.9 &2.94 \\
			S3-0 &0.05 & 4 &3 &9.8  &4.98 &-- & -- &-- &-- &-- \\
			D1 &0.05 & 3 &1 &10.3  &4.98 &0.05 & 3 &1 &10.3 &4.98 \\		\end{tabular}
	\end{ruledtabular}
	%	\footnotetext[1]{Here's the first, from Ref.~\onlinecite{feyn54}.}
	%	\footnotetext[2]{Here's the second.}
\end{table}

The mode composition of the simulated surfaces is estimated relying on the property of approximate orthogonality  of the wave modes. The mode amplitudes $b_n(x,t)$ and integrated
 over the simulation domain mode amplitudes $B_n(t)$ for given $k$ and mode number $n$ are calculated for a surface $\eta(x,y,t)$ as follows:
\begin{align} \label{ModeAmplitude}
		b_n(x,t)=\frac{\int{\eta Y_n dy}}{\int{Y_n^2 dy}},
		\quad B_n(t)= \left( \int{b_n^2 dx} \right)^{1/2}.
\end{align}
Orthogonal eigenmodes $Y_n(y)$ which are solutions of the Sturm--Liouville problem are used in this analysis.
Recall that since the BVP is determined by the longitudinal currier wavenumbers, in the simulations with different $k$, each $k$ yields specific set of transverse modes.
The function $b_n(x,t)$ corresponds to the solution of one-dimensional NLSE, $\text{Re}{\left( \psi_n(x,t) \exp{(i \omega_n t - ikx)} \right)} $. The integral mode amplitudes $B_n \ge0$ characterize the amount of energy held by the given transversal mode in the entire domain of simulation.

\begin{figure}[h]
	\includegraphics[width=8cm]{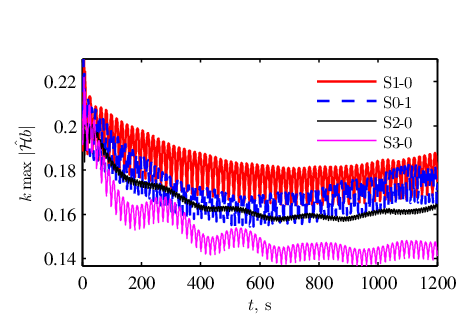} %proc1@ComparisonMaxModeAmplitudes@S1-0_vs_S0-1_vs_S2-0_vs_S3-0.eps}% Here is how to import EPS art
	\caption{\label{fig:TrueWaveSteepness} Maximum steepness of envelope solitons in the simulations of single solitons.}
\end{figure}

The evolution of amplitudes of the simulated single solitons is illustrated in Fig.~\ref{fig:TrueWaveSteepness}. The maximum wave steepness estimated as $k \max{|\hat{\cal{H}}b_n|}$ is plotted
 there, where the Hilbert transform $\hat{\cal{H}}$ with respect to the longitudinal coordinate is applied to capture the envelope of the mode function.
%Significant oscillations of the envelope remain due to the presence of nonlinear phase-locked wave harmonics.
One may see that while at the initial stage of the simulations the wave steepness may reach values exceeding $0.2$, the actual steepness of solitary groups for the time $t>400$~s lies approximately in the interval from $0.19$ down to $0.14$.

\subsection{\label{sec:SolitonsFundamentalMode}Interaction of solitons of the fundamental mode}

\begin{figure}[h]
	\includegraphics[width=8cm]{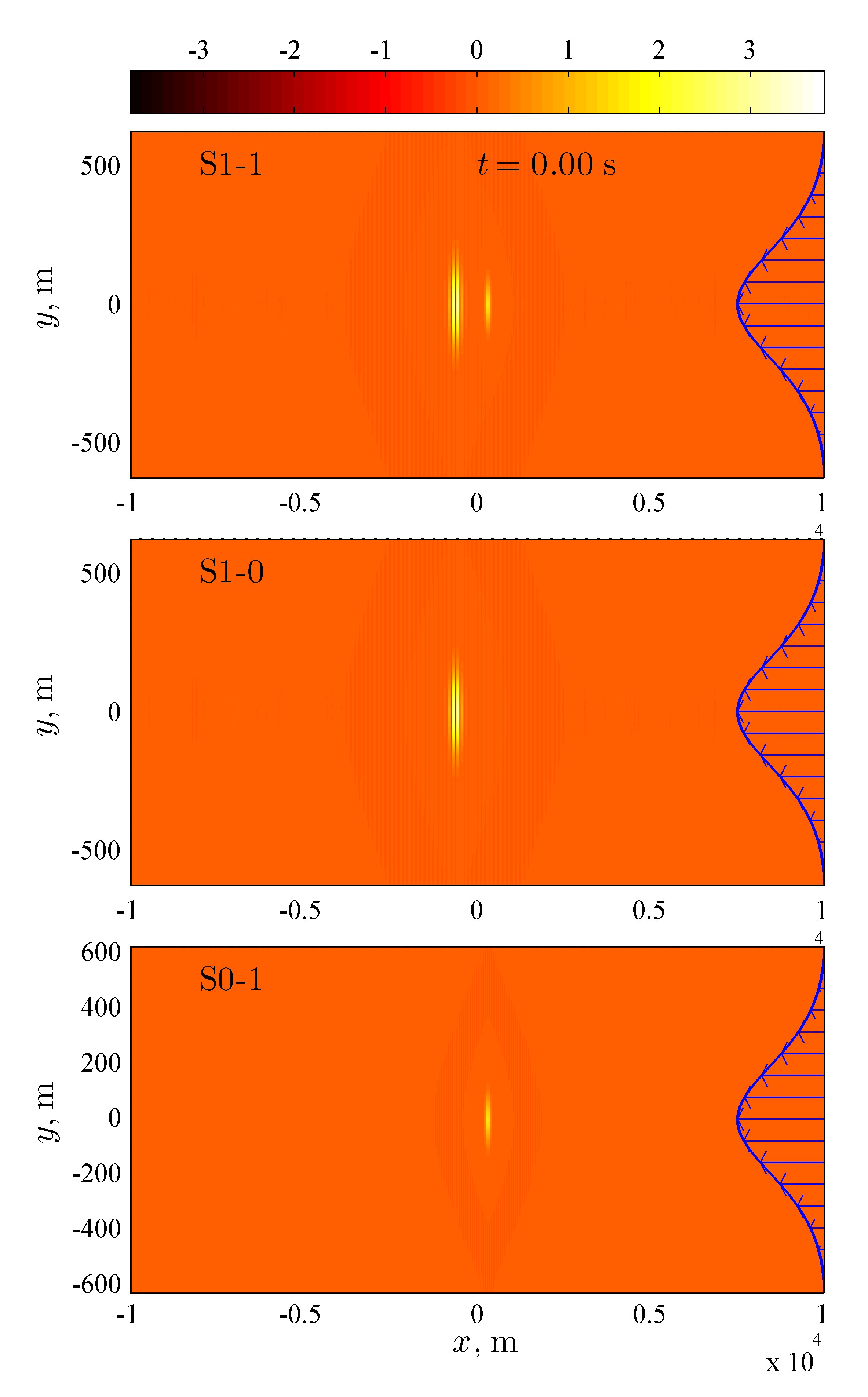} %proc1@ComparisonSurfaceRadiation@S1-1_vs_S1-0_vs_S0-1_0000.jpg}% Here is how to import EPS art
	\caption{\label{fig:Mode1&Mode1_0} Initial perturbations of the surfaces in the simulation of two interacting solitons S1-1 (top panel), just the first soliton S1-0 (middle panel) and the second soliton S0-1 (bottom panel). Waves propagate to the right; the current profile is given for the reference.}
\end{figure}

\begin{figure}[h]
	\includegraphics[width=8cm]{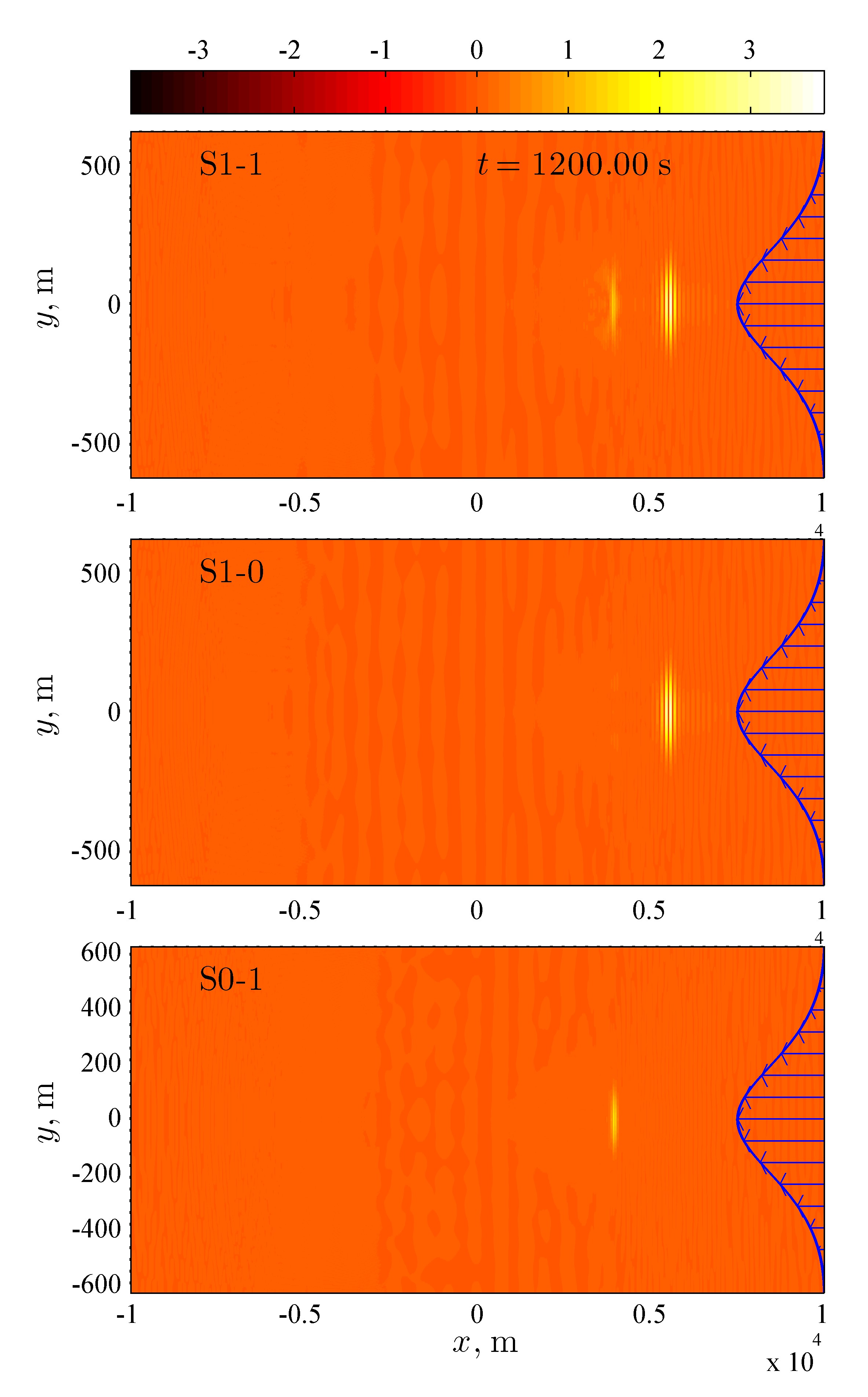} %proc1@ComparisonSurfaceRadiation@S1-1_vs_S1-0_vs_S0-1_2400.jpg} % Here is how to import EPS art
	\caption{\label{fig:Mode1&Mode1_1000} Surfaces at the end of the simulations with the initial conditions shown in Fig.~\ref{fig:Mode1&Mode1_0}. }
\end{figure}

\begin{figure}[h]
	\includegraphics[width=8cm]{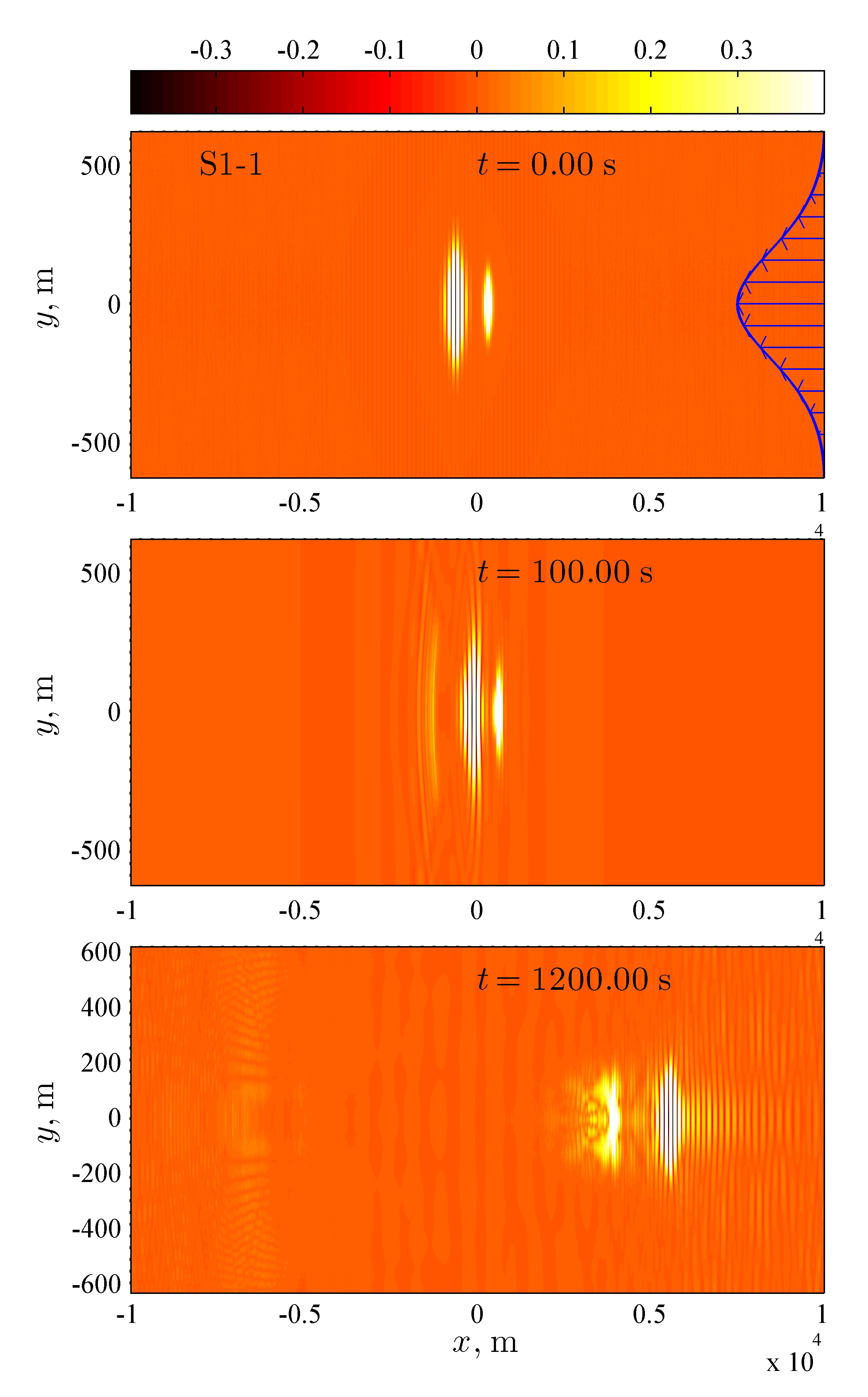} %proc1@ComparisonSurfaceRadiationOneExperiment@S1-1@0000_vs_0200_vs_2400_Zoom=10.jpg}% Here is how to import EPS art
	\caption{\label{fig:Radiation} Small-amplitude waves generated in the simulation S1-1. The surface displacements are 10 times magnified in amplitudes. The current profile is given in the top panel for the reference.}
\end{figure}

An example of the initial water surfaces at $t=0$ for the simulation of two solitons of the fundamental mode, $n_1=n_2=1$, which either interact or propagate alone, are given in Fig.~\ref{fig:Mode1&Mode1_0}. The surfaces at the end of the simulations $t=1200$~s are shown in Fig.~\ref{fig:Mode1&Mode1_1000}. The duration of simulation corresponds to almost $120$ periods of the first soliton carrier, and more than $150$ periods of the second soliton carrier.
The figures clearly show that the solitons  do survive the collision. After a deeper investigation, one can find that the shorter-wave soliton in the upper panel of Fig.~\ref{fig:Mode1&Mode1_1000} is in fact superimposed with a smaller-amplitude pattern of a complicated shape, which was emitted by the colliding solitons, see the low panel in Fig.~\ref{fig:Radiation}. The small amplitude long-scale radiation is also visible in all panels in Fig.~\ref{fig:Mode1&Mode1_1000}.

The radiation appears in the beginning of the simulation and is spreading later on in the entire simulation domain, as shown in Fig.~{\ref{fig:Radiation}}. In this figure the surface displacements are 10 times magnified in order to exhibit small-amplitude waves. The radiation consists of waves propagating to the left (which may be generated due to the imperfect balance between the surface displacement and the velocity potential of the initial condition, see the middle panel in Fig.~{\ref{fig:Radiation}}), and also forwarding waves of longer and shorter lengths. The length of the simulated domain $L_x = 6400 \pi$~m, what is about $20$~km, is enough to avoid the interaction between waves propagating to the left and to the right during the time of modeling, at least in the magnified amplitude scale of Fig.~{\ref{fig:Radiation}}.

The propagating waves may be residual waves due to the generation of phase-locked nonlinear bound waves, and also waves generated due to the inaccurate prescription of the envelope structure. The slightly inelastic collision of the soliton structures within the Euler equations causes new wave generation. The complicated patterns in the low panel of Fig.~\ref{fig:Radiation} may be associated with excitation of lateral modes with higher numbers. As was mentioned above, the low modes have very close velocities (see Fig.~\ref{fig:ModeVelocities}), therefore they can travel in combination for a long time.
As follows from  Fig.~{\ref{fig:Radiation}}, most of the radiation remains confined to the mainstream of the current.

\begin{figure}[h]
	\includegraphics[width=8cm]{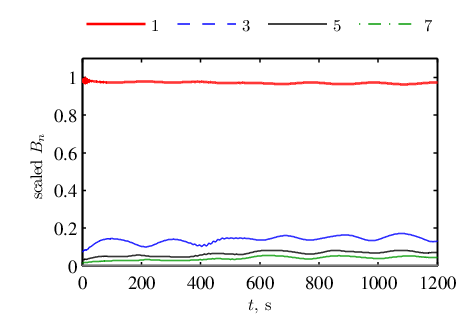} %mode_analys5_SturmLiouville@Run40@k0=0.05@ModeAmplitudesLargest.eps}% Here is how to import EPS art
	\caption{\label{fig:ModeAmplitudesMode1&Mode1} The Sturm--Liouville mode integral amplitudes $B_n(t)/B_1(0)$ for the carrier wavenumber $k=0.05$~rad/m in the experiment S1-1.}
\end{figure}

The evolution of integral mode amplitudes in the simulation of two solitons with different wave lengths which belong to fundamental modes is shown in Fig.~\ref{fig:ModeAmplitudesMode1&Mode1} for the carrier wavenumber $k=0.05$~rad/m. In the figure, the values of $B_n(t)$ are scaled by the maximum amplitude at the initial moment, which is $B_1(0)$ in this case. The most intense modes are plotted. The amplitudes of modes with even numbers are negligibly small, what is dictated by the symmetry of the simulated wave fields with respect to the plane $y=0$. The choice of the eigenfunctions $Y_n$ used for the calculation of the mode amplitudes according to (\ref{ModeAmplitude}), either from the Sturm--Liouville problem (\ref{BVP-SL}) or the nonlinear problem (\ref{BVP1D}), has very little effect on the amplitudes $B_n$ with significant values.

One can see from Fig.~\ref{fig:ModeAmplitudesMode1&Mode1} that besides the  first mode we wanted to study, there are also some other modes with small but non-zero amplitudes at the initial instant  $t=0$. This is due to the presence of the second soliton with the carrier wavenumber $k=0.1$~rad/m. In the simulations of single solitons, the amplitudes of other modes at $t=0$ are zeros. In the course of the evolution, some variations of the mode amplitudes are observed, as shown in Fig.~\ref{fig:ModeAmplitudesMode1&Mode1}, but they are of oscillatory type and are small in magnitude.
The period of beating between the first and the third mode $2\pi/(\omega_3-\omega_1)$ is estimated as about $188$~s and $194$~s according to the solution of the Sturm--Liouville problem and the nonlinear BVP respectively, which is close to the period of oscillations of $B_3(t)$ and also $B_1(t)$ shown in Fig.~\ref{fig:ModeAmplitudesMode1&Mode1} (the latter is difficult to discern in the scales of the figure). Small-amplitude fast oscillations of $B_1(t)$ at small $t$ is obviously due to the transition of the linear initial condition to the nonlinear wave.

Similar to  Fig.~\ref{fig:ModeAmplitudesMode1&Mode1} dependencies of mode amplitudes for the shorter carrier of the second soliton, $k=0.1$~rad/m, are not worthy of investigation, because of the mode overlaping with a longer and more energetic first soliton.

\subsection{\label{sec:SolitonsDifferentModes}Interaction of solitons of different transverse modes}

Collisions between solitons of different lateral modes have been found to occur in a qualitatively similar manner. In the examples of simulations we present in this subsection, the envelope soliton of shorter waves, $k_2=0.1$~rad/m, belongs to the fundamental mode, whereas the longer carrier wave soliton, $k_1=0.05$~rad/m, is configured according to the second mode (run S2-1) and the third mode (run S3-1). Note that the robustness of a single soliton of even higher, the 5-th mode has been already demonstrated in the numerical simulation in Ref.~\onlinecite{ShriraSlunyaev2014PRE}.

\begin{figure}[h]
	\includegraphics[width=8cm]{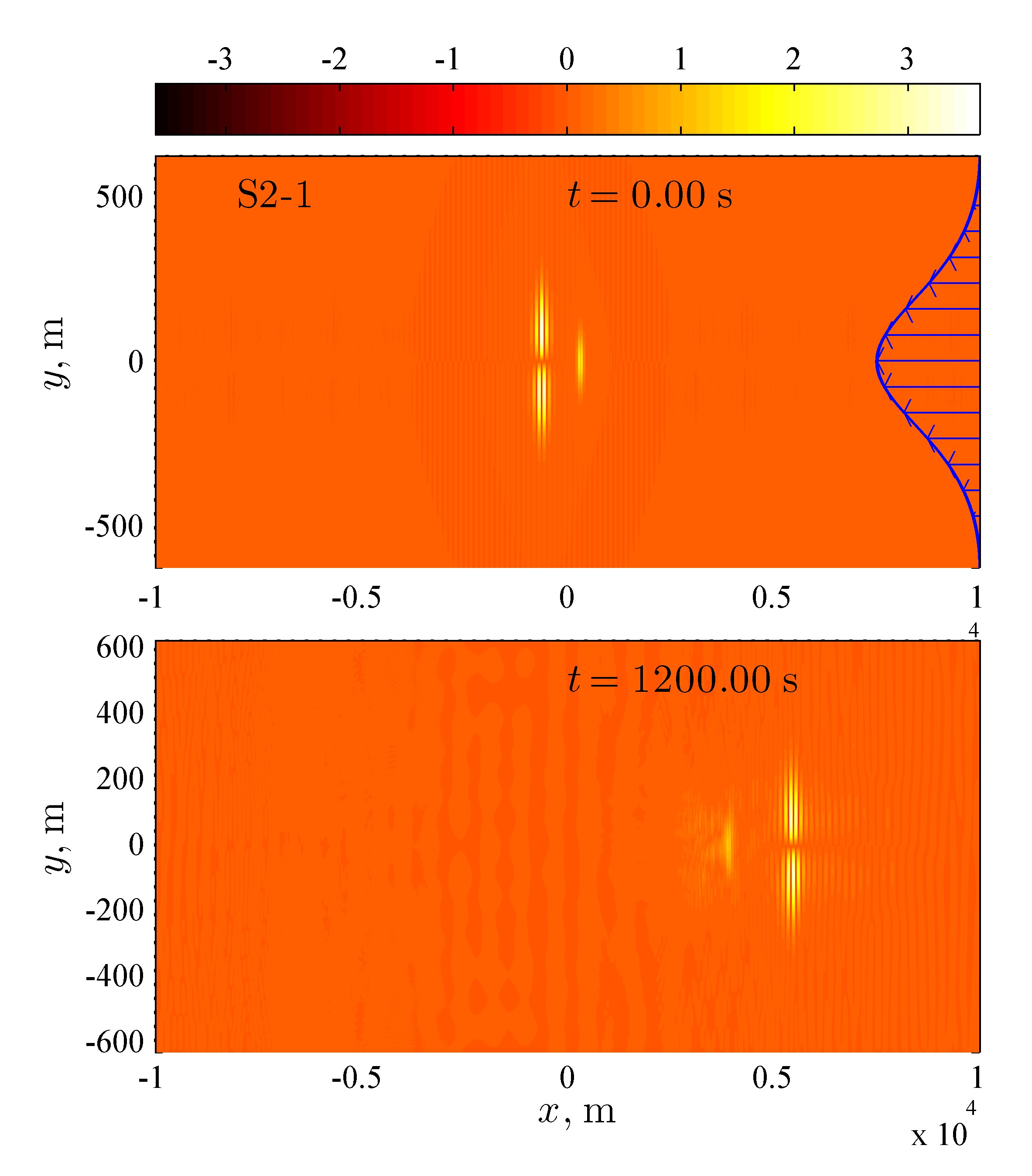} %proc1@ComparisonSurfaceRadiationOneExperiment@S2-1@0000_vs_2400_Zoom=1.1.jpg} % Here is how to import EPS art
	\caption{\label{fig:Mode2&Mode1} Interaction between envelope solitons of the second and the first modes (run S2-1): the initial condition (above) and the terminal moment  (below).}
\end{figure}

\begin{figure}[h]
	\includegraphics[width=8cm]{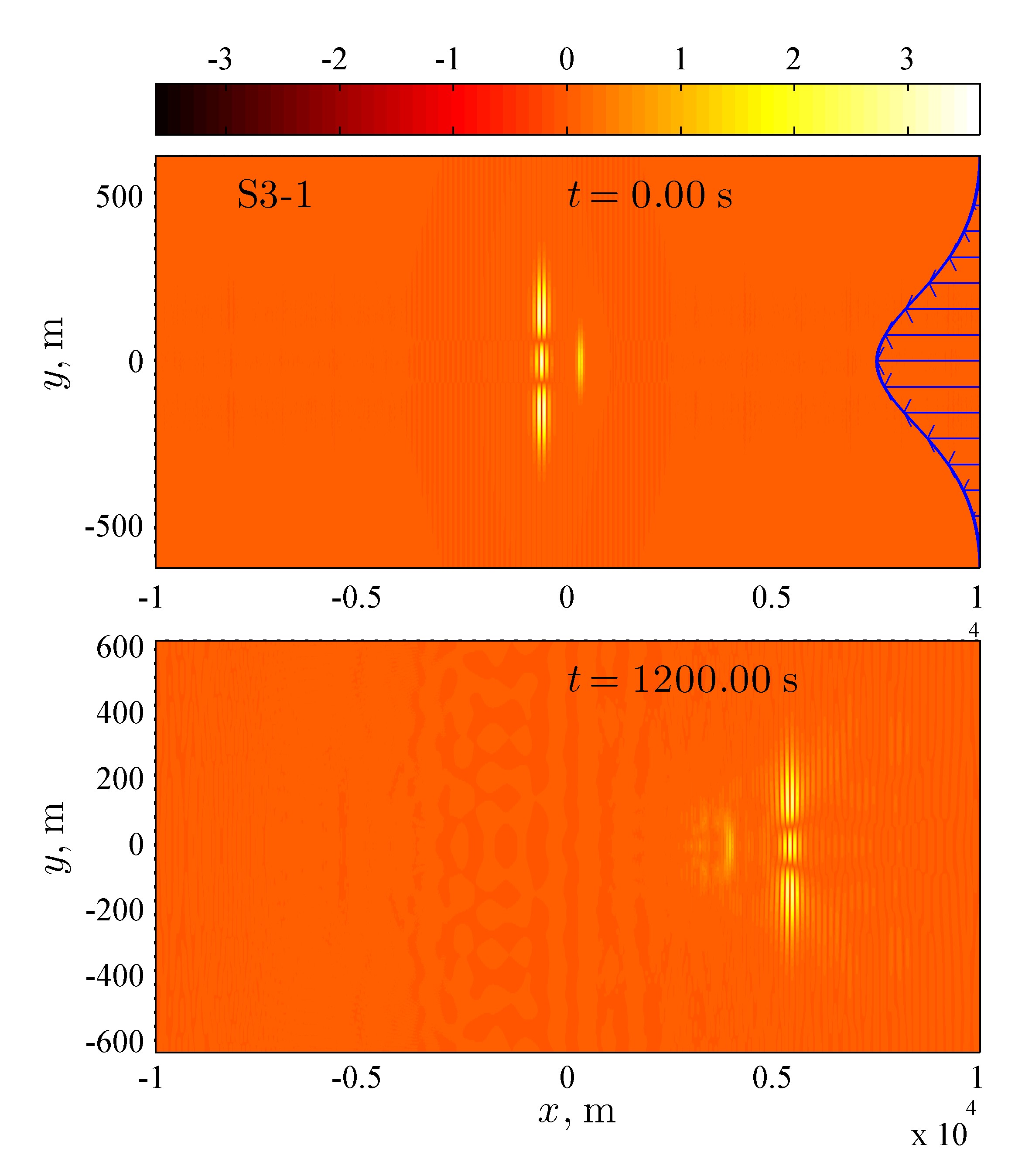} %proc1@ComparisonSurfaceRadiationOneExperiment@S3-1@0000_vs_2400_Zoom=1.1.jpg} % Here is how to import EPS art
	\caption{\label{fig:Mode3&Mode1} Interaction between envelope solitons of the third and the first modes (run S3-1): the initial condition (above) and the terminal moment (below).}
\end{figure}

The surfaces at the initial moment, $t=0$, and the instant when the simulations terminate at $t=1200$~s, are shown in Fig.~\ref{fig:Mode2&Mode1} and Fig.~\ref{fig:Mode3&Mode1} for the second and the third mode, respectively.
Figs.~\ref{fig:Mode2&Mode1} and \ref{fig:Mode3&Mode1} show that the envelope solitons recover after collisions in the both simulations, though the final solitary patterns look slightly less localized along the propagation direction, which corresponds to a reduction of their amplitudes in the course of the propagation and collision. According to the envelope soliton solution (\ref{OneSoliton}), smaller solitons are longer. The simulated solitons  preserve their specific lateral group structures prescribed by solutions of the BVP.  

\begin{figure}[h]
	\includegraphics[width=8cm]{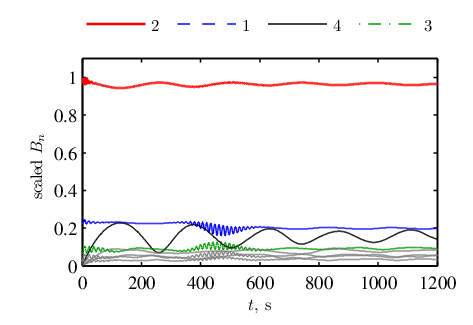} %mode_analys5_SturmLiouville@Run43@k0=0.05@ModeAmplitudesLargest.eps}% Here is how to import EPS art
	\caption{\label{fig:ModeAmplitudesMode2&Mode1}The Sturm--Liouville mode integral amplitudes $B_n(t)/B_2(0)$ for the carrier wavenumber $k=0.05$~rad/m in the experiment S2-1. The modes with even smaller energies are plotted by light grey lines.}
\end{figure}

\begin{figure}[h]
	\includegraphics[width=8cm]{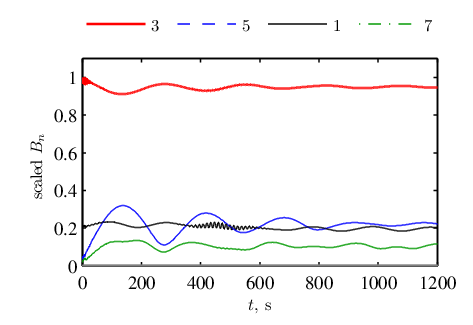} %mode_analys5_SturmLiouville@Run45@k0=0.05@ModeAmplitudesLargest.eps}% Here is how to import EPS art
	\caption{\label{fig:ModeAmplitudesMode3&Mode1} The Sturm--Liouville mode integral amplitudes $B_n(t)/B_3(0)$ for the carrier wavenumber $k=0.05$~rad/m in the experiment S3-1.}
\end{figure}

The integral mode amplitudes for these simulations are plotted in Fig.~\ref{fig:ModeAmplitudesMode2&Mode1} and Fig.~\ref{fig:ModeAmplitudesMode3&Mode1}.
Due to the overlap between modes of waves with different carrier wavenumbers, more than one mode amplitude is non-zero at $t=0$. The mode amplitudes exhibit quasi-periodic decaying oscillations in time with no clear trend to grow.
The periods of beating between the second and the fourth modes $2\pi/(\omega_4-\omega_2)$ and the third and the firth modes $2\pi/(\omega_5-\omega_3)$, which are  the most 
energetic ones  according to Fig.~\ref{fig:ModeAmplitudesMode2&Mode1} and Fig.~\ref{fig:ModeAmplitudesMode3&Mode1}, are about $209-219$~s and $235-248$~s respectively.
%(The solutions of the nonlinear BVP (\ref{BVP1D}) give a little bit longer values.)
The observed amplitude oscillations may be a result of the discrepancy between the solution in the  Sturm--Liouville approximation and the actual nonlinear wave modes.

A greater number of small-amplitude modes is excited in the situations when the interacting solitons belong to modes with different evenness, see the run S2-1 in Fig.~\ref{fig:ModeAmplitudesMode2&Mode1}.
Amplitudes of such 'parasitic' modes seem to increase with the soliton mode number (cf. Fig.~\ref{fig:ModeAmplitudesMode1&Mode1}, Fig.~\ref{fig:ModeAmplitudesMode2&Mode1} and Fig.~\ref{fig:ModeAmplitudesMode3&Mode1}), which is also observed in the simulations of single solitons S1-0, S2-0 and S3-0. The most likely interpretation of this observation is a larger discrepancy between the shapes of high transverse modes in the strongly nonlinear simulation and the approximate boundary value problem. Within the framework of the Euler equations, the effect of mode non-orthogonality likely becomes more noticeable, which  makes the mode amplitudes $b_n(x,t)$ and $B_n(t)$ a bit less meaningful.

\subsection{\label{sec:SolitonsMaxima}Extreme waves in the course of envelope soliton collisions}

\begin{figure*}[ht!]
	\includegraphics[width=16cm]{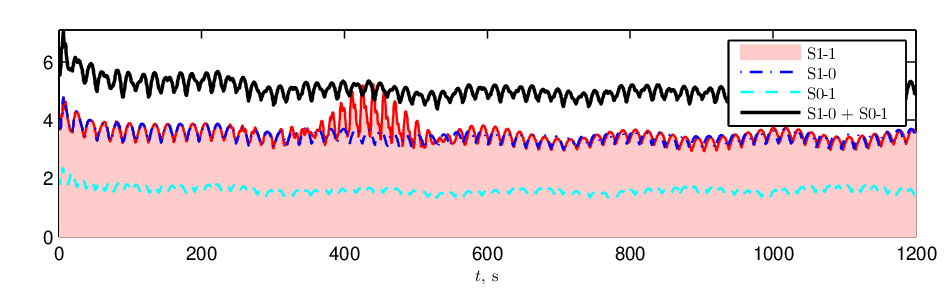}(a) \\% Here is how to import EPS art
	\includegraphics[width=16cm]{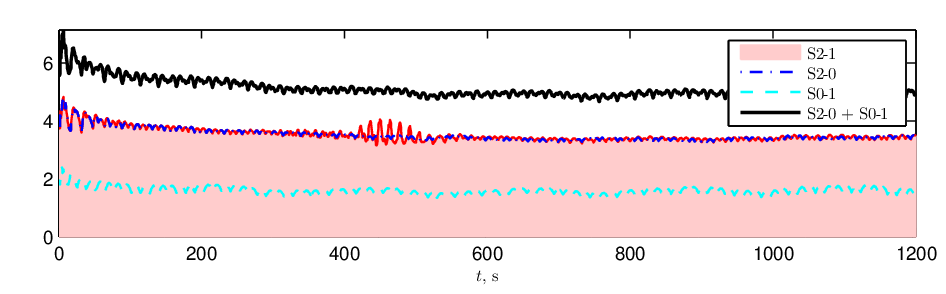}(b) \\% Here is how to import EPS art
	\includegraphics[width=16cm]{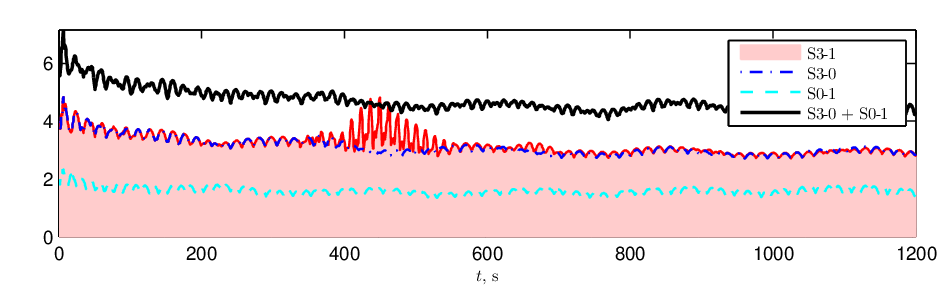}(c)% Here is how to import EPS art
	\caption{\label{fig:SolitonWaveAmplitudes} Evolution of the surface extremes $\max_{(x,y)}{|\eta(x,y,t)|}$ for the simulations of interacting solitons (shading with red curve above) and the simulations of independent solitons, against the sum of extremes for experiments with single solitons shown by the solid black curve. The interactions between solitons of the fundamental mode (a), second and fundamental mode (b) and third and fundamental mode (c) are shown.}
\end{figure*}

Extreme values of the surface displacement in the entire simulation domain per each time instant, are shown in Fig.~\ref{fig:SolitonWaveAmplitudes} for the three scenarios of interaction between pairs of solitons discussed above. The pink filling bounded from above by the red curve represents the simulation of interacting solitons, while the dash-dotted blue and broken cyan curves correspond to the simulations of independent solitons. The solid black curve denotes the  sum of the extreme displacements in the two simulations of single solitons. Thus, in every panel in Fig.~\ref{fig:SolitonWaveAmplitudes} the solid black curve represents the linear combination of two independent solitons, whereas the filling with the red upper edge describes the result of nonlinear interaction between the solitons of trapped waves.

One can see intense oscillations at the begining of the simulations caused by the nonlinear wave adjusting and formation of the 'true' strongly nonlinear patterns which correspond to stable envelope solitons of trapped waves.
Some fast and slow oscillations of the curves for wave extremes occur during the entire period of simulations. They appear due to the difference between the wave phase and group velocities and also due to interactions with radiated waves. Except for the time interval of collisions ($t=400...500$~s), the extremes in the simulations of interacting solitons generally follow the curves of extreme displacements in the simulations of single solitons with longer carrier (the blue dash-dotted curves).

During the collisions, the extreme displacements  reach the values of the linear superposition of two independent solitons or even exceed them in the situations of odd mode numbers, when the maxima of interacting soliton modes are located at $y=0$. The collision between waves of odd and even modes produces relatively smaller waves (Fig.~\ref{fig:SolitonWaveAmplitudes}~b) due to different lateral locations of the modes maxima. Within the NLSE, interacting envelope solitons can at best reach the sum of their amplitudes occurring in the independent evolution.

According to Figs.~\ref{fig:ModeAmplitudesMode1&Mode1}, \ref{fig:ModeAmplitudesMode2&Mode1}, \ref{fig:ModeAmplitudesMode3&Mode1} the mode integral amplitudes may  quickly oscillate during the stage of soliton collision, but this process does not lead to a noticeable redistribution of the energy between modes.

\section{\label{sec:DegenerateSolitons} Evolution of degenerate 2-soliton solution}

\begin{figure}[h]
	\includegraphics[width=8cm]{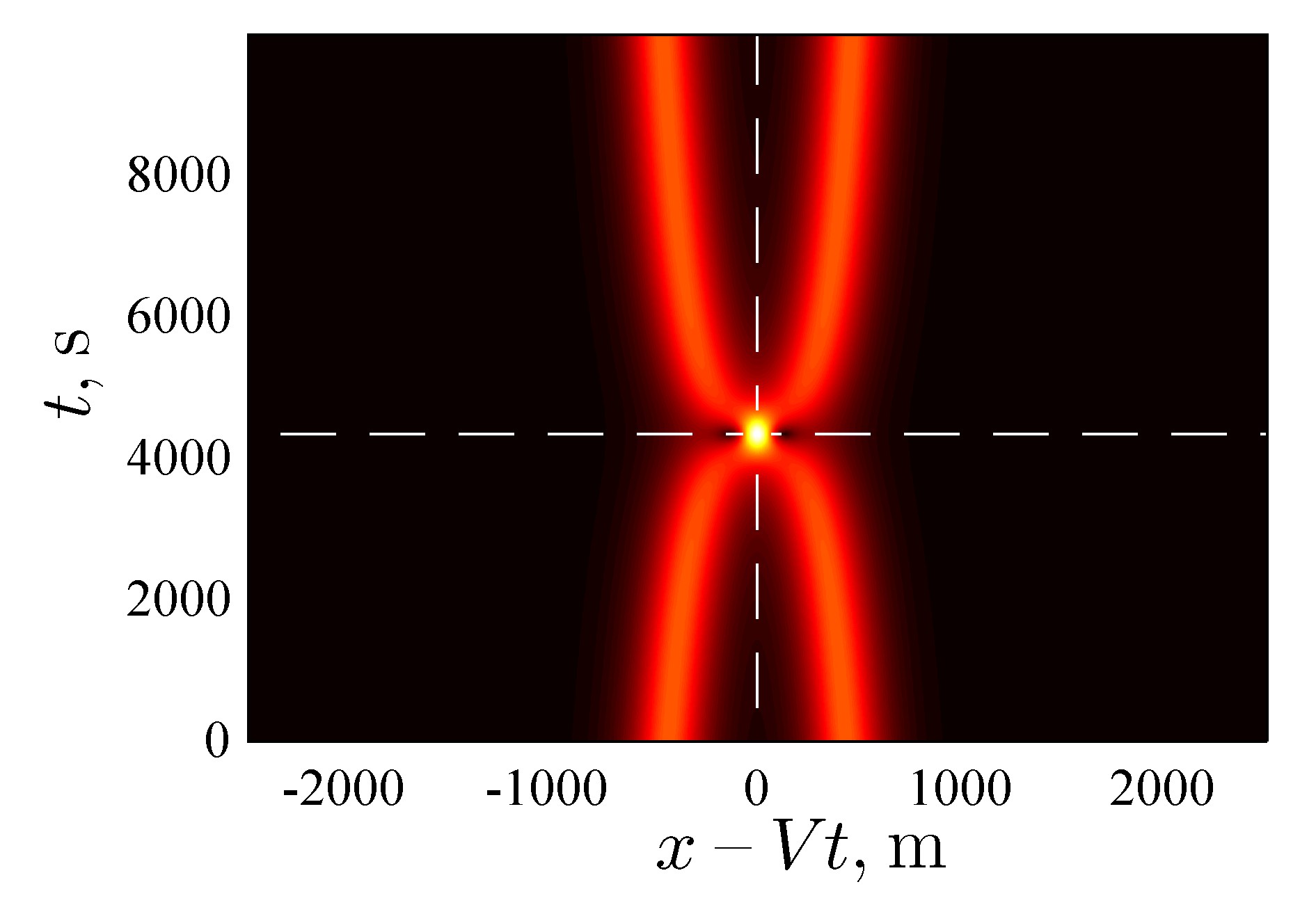}(a)\\
	\includegraphics[width=8cm]{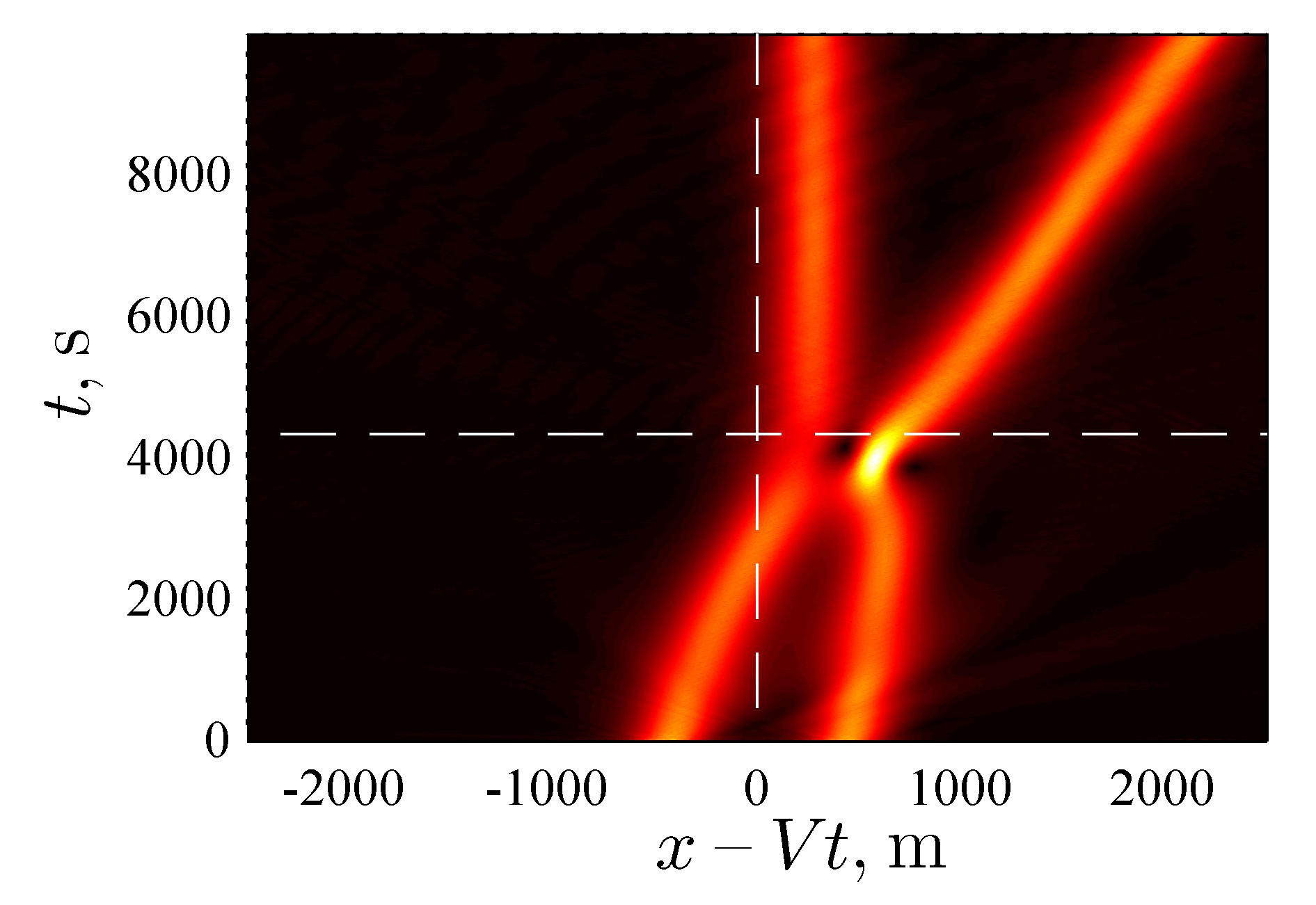}(b)
	\caption{\label{fig:DegenerateEnvelopeEvolution} Exact degenerate 2-soliton solution of the NLSE $|\psi_{\text{deg}}(x,t+T_f)|$ for $T_f=4343$~s, $k=0.05$~rad/m and the characteristic steepness $kA=0.15$ (a), and the envelope for function $b_1(x,t)$ in the following coordinate with $V\approx 5$~m/s in the numerical simulation D1 of the degenerate solution (b). According to the NLSE solution, the maximum wave should occur at $t=4343$~s, at the intersection of  the dashed lines.}
\end{figure}

Envelope solitons with different carrier wavelengths considered in Sec.~\ref{sec:2SolInteractions} which travel with substantially different speeds. The soliton velocity also depends on the mode shape. A collision of two solitons with close velocities takes very long time, its simulation would require resource demanding computations. If the initial condition is specified in the form of individual solitons located far from each other, the  simulation may be challenging in view of accumulating errors. 
To circumvent  this difficulty, the exact two-soliton solution of the NLSE may be used to determine the initial condition (as was done in Ref.~\onlinecite{Slunyaev2009} for modeling of unidirectional envelope solitons within the primitive equations of hydrodynamics). Two solitons with the same carrier wave, but with different amplitudes form a bound state (so-called bi-soliton\cite{Peregrine1983}), which periodically produces huge waves with the peak amplitude equal to the sum of amplitudes of the tangled solitons. An exact solution which describes a degenerate bi-soliton for the case when the soliton amplitudes coincide, is also known \cite{Peregrine1983,AkhmedievetalAnkiewicz1993}. Within this solution of the NLSE, two identical envelope solitons  experience attraction and approach each other, reduce the amplitudes before they merge, and then, for a short time, form an envelope with twice the amplitude of individual solitons. After that the solitons repulse, restore their shapes and drift infinitely far from each other, see the $x-t$ diagram of the solution in Fig.~\ref{fig:DegenerateEnvelopeEvolution}a.

The degenerate solution of the generalized NLSE  (\ref{NLS}) for the case when the nonlinear interaction  terms, except for the self-interaction, are neglected (i.e., all $\mu_j$ and $\nu_{npqr}$ are zeros)
may be presented in the following form:
\begin{align} \label{Degenerate}
	\psi_{\text{deg}}(x,t) = \nonumber \\
	= 4 \frac{a}{I_n} \frac{\xi \sinh{\xi}-\cosh{\xi} \left( 1+ \frac{i}{2} k^2 {a}^2 \omega_n t\right)}{\cosh{2 \xi}+1+2 \xi^2 + \frac{1}{2}  k^4 {a}^4 \omega_n^2 t^2} e^{i \frac{k^2 {a}^2}{4} \omega_n t}, \nonumber \\
	\text{where} \quad \xi=\sqrt{2} k^2a \left( x-Vt \right).
\end{align}
Though the expression looks rather complicated, it is the simplest form of a 2-soliton solution of the NLSE. In the asymptotics $t\to\pm\infty$ the solution splits into two distant envelope solitons (\ref{OneSoliton}) with amplitudes $A=a/I_n$. The maximum amplitude of the envelope is achieved at $t=0$ in the origin of the coordinate, $|\psi_{\text{deg}}(0,0)|=2A$. As discussed above, the solution represents the scenario with the longest possible interaction between solitons of one mode, which should be the  most favourable for highlighting inelastic features of the interaction.

The degenerate 2-soliton solution was simulated numerically and under laboratory conditions in Ref.~\onlinecite{Chabchoubetal2021}, where it was also  discussed in  the rogue wave context -- as a counterpart to the Peregrine breather without a background wave.

The numerical simulation of the degenerate 2-soliton solution (\ref{Degenerate}) for the fundamental mode of trapped waves, $n=1$, within the primitive equations of hydrodynamics is performed in this work for the same jet current (see Sec.~\ref{sec:CurrentInDNLS}). The initial perturbation is determined according to (\ref{InitialCondition}), where $\psi_{0,j}= \psi_{\text{deg}}(x,-T_f)$ and $j=1$. The carrier wavenumber is taken to be  $k=0.05$~rad/m, and the amplitude of separated solitons is $A = 3$~m, so that the characteristic wave steepness of a single soliton is $kA=0.15$, see the experiment D1 in Table~\ref{tab:SimulationParameters}. The focusing time $T_f=4343$~s is chosen to be identical (in terms of the dimensionless NLSE) to the conditions simulated in Ref.~\onlinecite{Chabchoubetal2021}. The solitons at the begining of the simulation just weakly overlap, see  Fig.~\ref{fig:DegenerateEnvelopeEvolution} for $t=0$ .

\begin{figure}[h]
	\includegraphics[width=7cm]{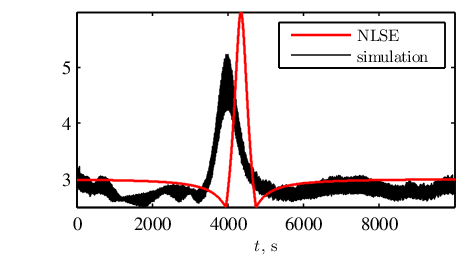} %plot_NLSsolution@Maximum@Run47.eps}
	\caption{\label{fig:DegenerateMaxima} Maximum of the analytic solution of the NLSE $|\psi_{\text{deg}}|$ and maximum of the  function  $|b_1|$ in the numerical simulation D1.}
\end{figure}

\begin{figure}[h]
	\includegraphics[width=7cm]{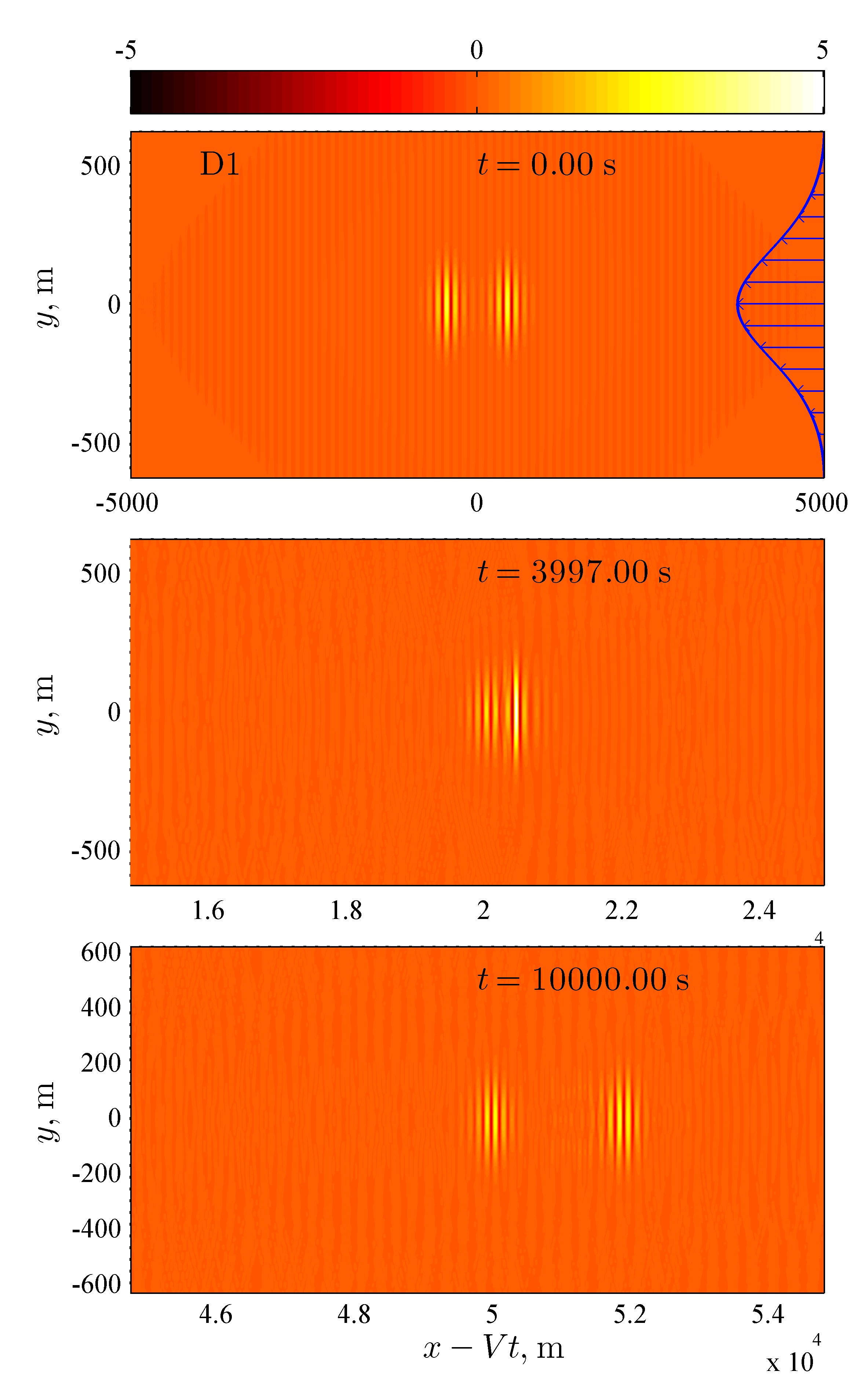} %proc1@ComparisonSurfaceRadiationOneExperiment@D1@0000_vs_3997_vs_10000.jpg}
	\caption{\label{fig:DegenerateSurfaces} Surfaces in the simulation D1 at the initial moment (top), at the moment of the maximal wave (middle) and at the end of the simulation (bottom) in the co-moving system of references.}
\end{figure}

The exact solution of the NLSE and the result of the simulation of the primitive equations are compared in Figs.~\ref{fig:DegenerateEnvelopeEvolution}, \ref{fig:DegenerateMaxima}. In order to represent the result of numerical solution, the function $b_1(x,t)$ is calculated according to (\ref{ModeAmplitude}), where the mode $Y_1(y)$ corresponds to the solution of the Sturm--Liouville problem. The envelope of $b_1$ is obtained at each instant using the Hilbert transform, $|\hat{\cal{H}} b_1|$. The mode velocity $V=V_1\approx 4.978$~m/s is calculated according to Eq.~($\ref{NLS}$).
The maximum surface displacements as functions of time according to the analytic solution of the NLSE, $\max_x{|\psi|}$, and estimated using the mode amplitude, $\max_{x}{|b_1|}$, are shown in Fig.~\ref{fig:DegenerateMaxima}; the latter behaves very similar to $\max_{(x,y)}{|\eta|}$ but is slightly larger in magnitude. The simulated water surfaces at the start of the simulation, at the end of the simulation, and also at the moment when the maximum surface displacement occurs, are shown in Fig.~\ref{fig:DegenerateSurfaces}.

The numerical simulation exhibits a picture qualitatively similar to the exact solution dynamics: the solitons approach each other and finally merge for some time; they produce strongly amplified waves at approximately the time $T_f$ predicted by the exact NLSE solution, and then split in two stable isolated patterns. According to Fig.~\ref{fig:DegenerateMaxima}, the maximum wave amplitude in the end of the simulation is very close to that of the initial condition. No significant radiation or external waves are present in Fig.~\ref{fig:DegenerateEnvelopeEvolution}b; only small-amplitude waves propagating away from the simulated wave structure can be found in Fig.~\ref{fig:DegenerateSurfaces}.

However, the symmetry of the spatio-temporal diagram (Fig.~\ref{fig:DegenerateEnvelopeEvolution}a) brakes down when simulated within the strongly nonlinear framework (Fig.~\ref{fig:DegenerateEnvelopeEvolution}b).  Compared to the mode velocity $V$, the simulated solitons move slightly faster. In the course of collision, the leading soliton acquires more energy and starts moving even faster after the collision. Extreme waves with amplitudes slightly exceeding $5$~m are generated in the course of the collision instead of $6$~m predicted by the analytic solution; the focusing event occurs a little earlier.

\begin{figure}[h]
	\includegraphics[width=8cm]{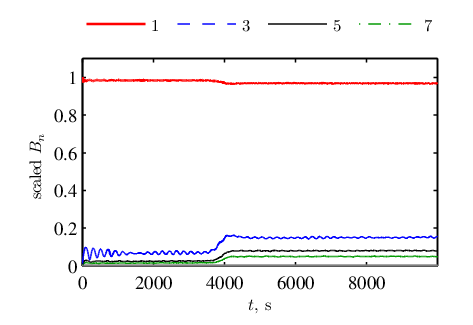} %mode_analys5_SturmLiouville@Run47@k0=0.05@ModeAmplitudesLargest.eps}
	\caption{\label{fig:DegenerateModeAmplitudes} Sturm--Liouville mode integral amplitudes $B_n(t)/B_1(0)$ for the carrier wavenumber $k=0.05$~rad/m in the experiment D1.}
\end{figure}

The inelastic aspects of the soliton interaction in the degenerate situation are readily seen in Fig.~\ref{fig:DegenerateModeAmplitudes}, where the integral mode amplitudes are plotted versus time. It is clear that the collision of the solitons with generation of steeper waves at $t\approx4000$~s results in a noticeable increase of amplitudes of other modes, which do not relax after the collision episode. At the same time, the related decrease of the amplitude of the first mode is fairly small. The simulation lasts for almost $1~000$ wave periods, when the solitons with the carrier wave length $125$~m cover the distance of about 50 kilometers. Therefore, one may conclude that the solitons are robust, and the longest possible interaction between envelope solitons of trapped by the jet current waves occurs almost elastically.

%We also conducted another simulation of the degenerate soliton solution with twice shorter waves, $k=0.1$~rad/m, which all other parameters were the same as in D1. In this simulation the overall wave dynamics reproduced the one in D1, but the maximum wave compared to the analytic solution was somewhat smaller than in Fig.~\ref{fig:DegenerateMaxima}, and the amount of energy transferred from the first mode to the others was more significant than in Fig.~\ref{fig:DegenerateModeAmplitudes}. Hence, the case with $k=0.1$~rad/m was found to demonstrate a less elastic interaction.

\section{\label{sec:Conclusion} Conclusions}

Envelope solitons of water waves trapped by opposing jet currents realize a nearly unique situation where surface wave groups in deep water may be localized in both horizontal directions \cite{ShriraSlunyaev2014PRE}. Recall that the 2D nonlinear Schr\"odinger equation for modulations of deep-water waves on the two-dimensional surface does not possess stable soliton-type solutions localized in the two spatial dimensions, while the planar envelope solitons which are straightforward to construct by a simple change of variables from the solutions of the 1D NLSE, are known to be  transversally unstable \cite{ZakharovRubenchik1974}.

In the present work we demonstrate by means of the direct numerical simulation of the Euler equations  that envelope solitons of trapped waves (at least of low modes) are robust enough to exhibit, 
although only approximately, the key properties of the classic solitons -- localized solutions of integrable equations, such as: very long propagation preserving their structure and  energy, and,
 to a great extent, elastic pairwise collisions. In our study the soliton-type nonlinear groups were generated using approximate initial solutions, thus the fact that the  solitons of trapped waves 
 emerge as result of such excitations shows that  these patterns are attractors. %,   in the sense of the nonlinear dynamics theory.
The results of the numerical simulation also  show that the nonlinear inter-mode exchange during evolution of trapped  nonlinear waves on jet currents is insignificant (at least in some class of wave conditions), which makes the mode 
theory developed in Ref.~\onlinecite{ShriraSlunyaev2014JFM} effective. In particular, this finding justifies, to the leading order, consideration of dynamics of different modes as independent. Although emphasizing  the overall qualitative agreement, note that some quantitative discrepancies between the approximate analytic solutions obtained under the assumptions of weak nonlinearity, weak dispersion, and weak current compared to the numerical simulation of the Euler equations have been also found. In particular, interactions between solitons of trapped waves are not fully elastic; new waves of small amplitude are emitted.

The presented demonstration of qualitative validity of weakly-nonlinear soliton solutions up to the range of unexpectedly strong nonlinearity, up to the characteristic wave steepness $kA\approx0.2$, is in accordance with the recent numerical and laboratory simulations of strongly nonlinear counterparts of exact solutions of approximate modulation equations, see Refs.~\onlinecite{Slunyaev2009,SlunyaevShrira2013,Slunyaevetal2013PRE,Slunyaevetal2017,Chabchoubetal2021,Heetal2022}. We note, that we simulated an example of the interaction between even steeper solitons of the fundamental mode with $kA=0.25$ (not described in this work), which, again, demonstrated almost elastic collision.

The  simulations reported in this paper were carried out with the nonlinear parameter of the HOSM numerical code $M=5$ (which confines the consideration to  up to 6-wave interactions). They were also repeated in the series of experiments with a lower order of nonlinearity, $M=3$, which is sufficient to describe up to 4-wave interactions only. Those experiments exhibited very similar dynamics.

Coupling of solitons which originally propagate with close velocities is common  of weakly non-integrable wave dynamics. In particular, formation of a transient bound state in the strongly nonlinear simulation of degenerate 2-soliton NLSE solution of unidirectional waves was observed in Ref.~\onlinecite{Chabchoubetal2021} for the characteristic wave steepness $kA=0.15$. In contrast, in the case of waves trapped by the jet current, the solitons' coupling did not occur in the conducted simulation of degenerate solution with the same steepness and the same initial distance between the solitons.

One of the limitations of the adopted approach is the assumption of potentiality of the wave motion, which, although common,  is, strictly speaking, not true. However, a rough estimate of possible vorticity contribution to dynamics of waves suggests it to be very weak, so that at the time scales of formation and evolution of envelope solitons we are interested in,   it can be neglected. A detailed analysis of the role of vorticity of water wave field in the dynamics of trapped waves although being an interesting issue goes beyond the scopes of this work and requires a dedicated study. Some discussion of the effect of vorticity on the NLSE theory for waves upon horizontally inhomogeneous weak current may be found in Ref.~\onlinecite{HjelmervikTrulsen2009}.

The demonstrated effects of robustness of spatially localized intense wave groups trapped by opposite jet currents with respect to collisions  is expected to  be important in the context of 
anomalous oceanic waves (rogue waves). Whereas in still water the effect of finite wave directionality greatly reduces the strength of the modulational instability and the chance of envelope 
quasi-solitons to propagate for a relatively long period of time,  solitons of trapped waves can preserve their structure remarkably long. Hence, they can increase the 
probability of high waves with no restriction on the width of the wave angular spectrum. This essentially nonlinear effect might be a factor in the observed high likelihood of rogue wave occurrences 
on currents. The robustness of solitons of trapped waves opens the possibility of investigating of specific physical mechanisms caused by the account of inhomogeneity of the currents. 
Both the  rapid and adiabatically slow transformation of envelope solitons of trapped waves can amplify them and lead to even more extreme wave events.
However, these issues are beyond the scope of this work and will be considered elsewhere. 
 %The modal wave theory can provide an efficient theoretical description.

\begin{acknowledgments}
The research was supported by the Russian Science Foundation, grant No 22–17-00153.
\end{acknowledgments}

\section*{Data Availability Statement}

Data available on request from the authors.

% The \nocite command causes all entries in a bibliography to be printed out
% whether or not they are actually referenced in the text. This is appropriate
% for the sample file to show the different styles of references, but authors
% most likely will not want to use it.
%\nocite{*}

\bibliography{ReferenceBase}% Produces the bibliography via BibTeX.

%aipnum4-2.bst 2019-01-14 (MD) hand-edited version of apsrev4-1.bst
%Control: key (0)
%Control: author (8) initials jnrlst
%Control: editor formatted (1) identically to author
%Control: production of article title (0) allowed
%Control: page (1) range
%Control: year (1) truncated
%Control: production of eprint (0) enabled
\providecommand{\noopsort}[1]{}\providecommand{\singleletter}[1]{#1}%
\begin{thebibliography}{30}%
\makeatletter
\providecommand \@ifxundefined [1]{%
 \@ifx{#1\undefined}
}%
\providecommand \@ifnum [1]{%
 \ifnum #1\expandafter \@firstoftwo
 \else \expandafter \@secondoftwo
 \fi
}%
\providecommand \@ifx [1]{%
 \ifx #1\expandafter \@firstoftwo
 \else \expandafter \@secondoftwo
 \fi
}%
\providecommand \natexlab [1]{#1}%
\providecommand \enquote  [1]{``#1''}%
\providecommand \bibnamefont  [1]{#1}%
\providecommand \bibfnamefont [1]{#1}%
\providecommand \citenamefont [1]{#1}%
\providecommand \href@noop [0]{\@secondoftwo}%
\providecommand \href [0]{\begingroup \@sanitize@url \@href}%
\providecommand \@href[1]{\@@startlink{#1}\@@href}%
\providecommand \@@href[1]{\endgroup#1\@@endlink}%
\providecommand \@sanitize@url [0]{\catcode `\\12\catcode `\$12\catcode
  `\&12\catcode `\#12\catcode `\^12\catcode `\_12\catcode `\%12\relax}%
\providecommand \@@startlink[1]{}%
\providecommand \@@endlink[0]{}%
\providecommand \url  [0]{\begingroup\@sanitize@url \@url }%
\providecommand \@url [1]{\endgroup\@href {#1}{\urlprefix }}%
\providecommand \urlprefix  [0]{URL }%
\providecommand \Eprint [0]{\href }%
\providecommand \doibase [0]{https://doi.org/}%
\providecommand \selectlanguage [0]{\@gobble}%
\providecommand \bibinfo  [0]{\@secondoftwo}%
\providecommand \bibfield  [0]{\@secondoftwo}%
\providecommand \translation [1]{[#1]}%
\providecommand \BibitemOpen [0]{}%
\providecommand \bibitemStop [0]{}%
\providecommand \bibitemNoStop [0]{.\EOS\space}%
\providecommand \EOS [0]{\spacefactor3000\relax}%
\providecommand \BibitemShut  [1]{\csname bibitem#1\endcsname}%
\let\auto@bib@innerbib\@empty
%</preamble>
\bibitem [{\citenamefont {Shrira}\ and\ \citenamefont
  {Slunyaev}(2014{\natexlab{a}})}]{ShriraSlunyaev2014PRE}%
  \BibitemOpen
  \bibfield  {author} {\bibinfo {author} {\bibfnamefont {V.}~\bibnamefont
  {Shrira}}\ and\ \bibinfo {author} {\bibfnamefont {A.}~\bibnamefont
  {Slunyaev}},\ }\bibfield  {title} {\enquote {\bibinfo {title} {Nonlinear
  dynamics of trapped waves on jet currents and rogue waves},}\ }\href@noop {}
  {\bibfield  {journal} {\bibinfo  {journal} {Phys. Rev. E}\ }\textbf {\bibinfo
  {volume} {89}},\ \bibinfo {pages} {041002(R)} (\bibinfo {year}
  {2014}{\natexlab{a}})}\BibitemShut {NoStop}%
\bibitem [{\citenamefont {Mallory}(1974)}]{Mallory1974}%
  \BibitemOpen
  \bibfield  {author} {\bibinfo {author} {\bibfnamefont {J.}~\bibnamefont
  {Mallory}},\ }\bibfield  {title} {\enquote {\bibinfo {title} {Abnormal waves
  on the south east coast of south africa},}\ }\href@noop {} {\bibfield
  {journal} {\bibinfo  {journal} {The International hydrographic review}\
  }\textbf {\bibinfo {volume} {51}},\ \bibinfo {pages} {99--129} (\bibinfo
  {year} {1974})}\BibitemShut {NoStop}%
\bibitem [{\citenamefont {Kharif}, \citenamefont {Pelinovsky},\ and\
  \citenamefont {Slunyaev}(2009)}]{Kharifetal2009}%
  \BibitemOpen
  \bibfield  {author} {\bibinfo {author} {\bibfnamefont {C.}~\bibnamefont
  {Kharif}}, \bibinfo {author} {\bibfnamefont {E.}~\bibnamefont {Pelinovsky}},\
  and\ \bibinfo {author} {\bibfnamefont {A.}~\bibnamefont {Slunyaev}},\
  }\href@noop {} {\emph {\bibinfo {title} {Rogue Waves in the Ocean}}}\
  (\bibinfo  {publisher} {Springer-Verlag: Berlin Heidelberg},\ \bibinfo {year}
  {2009})\BibitemShut {NoStop}%
\bibitem [{\citenamefont {Peregrine}(1976)}]{Peregrine1976}%
  \BibitemOpen
  \bibfield  {author} {\bibinfo {author} {\bibfnamefont {D.}~\bibnamefont
  {Peregrine}},\ }\bibfield  {title} {\enquote {\bibinfo {title} {Interaction
  of water waves and currents},}\ }\href@noop {} {\bibfield  {journal}
  {\bibinfo  {journal} {Adv. Appl. Mech.}\ }\textbf {\bibinfo {volume} {16}},\
  \bibinfo {pages} {9--117} (\bibinfo {year} {1976})}\BibitemShut {NoStop}%
\bibitem [{\citenamefont {Smith}(1976)}]{Smith1976}%
  \BibitemOpen
  \bibfield  {author} {\bibinfo {author} {\bibfnamefont {R.}~\bibnamefont
  {Smith}},\ }\bibfield  {title} {\enquote {\bibinfo {title} {Giant waves},}\
  }\href@noop {} {\bibfield  {journal} {\bibinfo  {journal} {J. Fluid Mech.}\
  }\textbf {\bibinfo {volume} {77}},\ \bibinfo {pages} {417--431} (\bibinfo
  {year} {1976})}\BibitemShut {NoStop}%
\bibitem [{\citenamefont {White}\ and\ \citenamefont
  {Fornberg}(1998)}]{WhiteFornberg1998}%
  \BibitemOpen
  \bibfield  {author} {\bibinfo {author} {\bibfnamefont {B.}~\bibnamefont
  {White}}\ and\ \bibinfo {author} {\bibfnamefont {B.}~\bibnamefont
  {Fornberg}},\ }\bibfield  {title} {\enquote {\bibinfo {title} {On the chance
  of freak waves at the sea},}\ }\href@noop {} {\bibfield  {journal} {\bibinfo
  {journal} {J. Fluid Mech.}\ }\textbf {\bibinfo {volume} {255}},\ \bibinfo
  {pages} {113--138} (\bibinfo {year} {1998})}\BibitemShut {NoStop}%
\bibitem [{\citenamefont {Lavrenov}(1998)}]{Lavrenov1998}%
  \BibitemOpen
  \bibfield  {author} {\bibinfo {author} {\bibfnamefont {I.}~\bibnamefont
  {Lavrenov}},\ }\bibfield  {title} {\enquote {\bibinfo {title} {The wave
  energy concentration at the agulhas current of south africa},}\ }\href@noop
  {} {\bibfield  {journal} {\bibinfo  {journal} {Nat. Hazards.}\ }\textbf
  {\bibinfo {volume} {17}},\ \bibinfo {pages} {117--127} (\bibinfo {year}
  {1998})}\BibitemShut {NoStop}%
\bibitem [{\citenamefont {Hjelmervik}\ and\ \citenamefont
  {Trulsen}(2009)}]{HjelmervikTrulsen2009}%
  \BibitemOpen
  \bibfield  {author} {\bibinfo {author} {\bibfnamefont {K.}~\bibnamefont
  {Hjelmervik}}\ and\ \bibinfo {author} {\bibfnamefont {K.}~\bibnamefont
  {Trulsen}},\ }\bibfield  {title} {\enquote {\bibinfo {title} {Freak wave
  statistics on collinear currents},}\ }\href@noop {} {\bibfield  {journal}
  {\bibinfo  {journal} {J. Fluid. Mech.}\ }\textbf {\bibinfo {volume} {637}},\
  \bibinfo {pages} {267--284} (\bibinfo {year} {2009})}\BibitemShut {NoStop}%
\bibitem [{\citenamefont {Janssen}\ and\ \citenamefont
  {Herbers}(2009)}]{JanssenHerbers2009}%
  \BibitemOpen
  \bibfield  {author} {\bibinfo {author} {\bibfnamefont {T.}~\bibnamefont
  {Janssen}}\ and\ \bibinfo {author} {\bibfnamefont {T.}~\bibnamefont
  {Herbers}},\ }\bibfield  {title} {\enquote {\bibinfo {title} {Nonlinear wave
  statistics in a focal zone},}\ }\href@noop {} {\bibfield  {journal} {\bibinfo
   {journal} {J. Phys. Oceanogr.}\ }\textbf {\bibinfo {volume} {39}},\ \bibinfo
  {pages} {1948--1964} (\bibinfo {year} {2009})}\BibitemShut {NoStop}%
\bibitem [{\citenamefont {Onorato}, \citenamefont {Proment},\ and\
  \citenamefont {Toffoli}(2011)}]{Onoratoetal2011}%
  \BibitemOpen
  \bibfield  {author} {\bibinfo {author} {\bibfnamefont {M.}~\bibnamefont
  {Onorato}}, \bibinfo {author} {\bibfnamefont {D.}~\bibnamefont {Proment}},\
  and\ \bibinfo {author} {\bibfnamefont {A.}~\bibnamefont {Toffoli}},\
  }\bibfield  {title} {\enquote {\bibinfo {title} {Triggering rogue waves in
  opposing currents},}\ }\href@noop {} {\bibfield  {journal} {\bibinfo
  {journal} {Phys. Rev. Lett.}\ }\textbf {\bibinfo {volume} {107}},\ \bibinfo
  {pages} {184502} (\bibinfo {year} {2011})}\BibitemShut {NoStop}%
\bibitem [{\citenamefont {Ruban}(2012)}]{Ruban2012}%
  \BibitemOpen
  \bibfield  {author} {\bibinfo {author} {\bibfnamefont {V.}~\bibnamefont
  {Ruban}},\ }\bibfield  {title} {\enquote {\bibinfo {title} {On the nonlinear
  schr{\"o}dinger equation for waves on a nonuniform current,},}\ }\href@noop
  {} {\bibfield  {journal} {\bibinfo  {journal} {JETP. Lett.}\ }\textbf
  {\bibinfo {volume} {95}},\ \bibinfo {pages} {486--491} (\bibinfo {year}
  {2012})}\BibitemShut {NoStop}%
\bibitem [{\citenamefont {Shrira}\ and\ \citenamefont
  {Slunyaev}(2014{\natexlab{b}})}]{ShriraSlunyaev2014JFM}%
  \BibitemOpen
  \bibfield  {author} {\bibinfo {author} {\bibfnamefont {V.}~\bibnamefont
  {Shrira}}\ and\ \bibinfo {author} {\bibfnamefont {A.}~\bibnamefont
  {Slunyaev}},\ }\bibfield  {title} {\enquote {\bibinfo {title} {Trapped waves
  on jet currents: asymptotic modal approach},}\ }\href@noop {} {\bibfield
  {journal} {\bibinfo  {journal} {J. Fluid Mech.}\ }\textbf {\bibinfo {volume}
  {738}},\ \bibinfo {pages} {65--104} (\bibinfo {year}
  {2014}{\natexlab{b}})}\BibitemShut {NoStop}%
\bibitem [{\citenamefont {Akhmediev}, \citenamefont {Soto-Crespo},\ and\
  \citenamefont {Devine}(2016)}]{Akhmedievetal2016}%
  \BibitemOpen
  \bibfield  {author} {\bibinfo {author} {\bibfnamefont {N.}~\bibnamefont
  {Akhmediev}}, \bibinfo {author} {\bibfnamefont {J.~M.}\ \bibnamefont
  {Soto-Crespo}},\ and\ \bibinfo {author} {\bibfnamefont {N.}~\bibnamefont
  {Devine}},\ }\bibfield  {title} {\enquote {\bibinfo {title} {Breather
  turbulence versus soliton turbulence: Rogue waves, probability density
  functions, and spectral features},}\ }\href@noop {} {\bibfield  {journal}
  {\bibinfo  {journal} {Phys. Rev. E}\ }\textbf {\bibinfo {volume} {94}},\
  \bibinfo {pages} {022212} (\bibinfo {year} {2016})}\BibitemShut {NoStop}%
\bibitem [{\citenamefont {Kachulin}, \citenamefont {Dyachenko},\ and\
  \citenamefont {Dremov}(2020)}]{Kachulinetal2020}%
  \BibitemOpen
  \bibfield  {author} {\bibinfo {author} {\bibfnamefont {D.}~\bibnamefont
  {Kachulin}}, \bibinfo {author} {\bibfnamefont {A.}~\bibnamefont
  {Dyachenko}},\ and\ \bibinfo {author} {\bibfnamefont {S.}~\bibnamefont
  {Dremov}},\ }\bibfield  {title} {\enquote {\bibinfo {title} {Multiple soliton
  interactions on the surface of deep water},}\ }\href@noop {} {\bibfield
  {journal} {\bibinfo  {journal} {Fluids}\ }\textbf {\bibinfo {volume} {5}},\
  \bibinfo {pages} {65} (\bibinfo {year} {2020})}\BibitemShut {NoStop}%
\bibitem [{\citenamefont {Suret}\ \emph {et~al.}(2020)\citenamefont {Suret},
  \citenamefont {Tikan}, \citenamefont {Bonnefoy}, \citenamefont {Copie},
  \citenamefont {Ducrozet}, \citenamefont {Gelash}, \citenamefont
  {Prabhudesai}, \citenamefont {Michel}, \citenamefont {Cazaubiel},
  \citenamefont {Falcon}, \citenamefont {El},\ and\ \citenamefont
  {Randoux}}]{Suretetal2020}%
  \BibitemOpen
  \bibfield  {author} {\bibinfo {author} {\bibfnamefont {P.}~\bibnamefont
  {Suret}}, \bibinfo {author} {\bibfnamefont {A.}~\bibnamefont {Tikan}},
  \bibinfo {author} {\bibfnamefont {F.}~\bibnamefont {Bonnefoy}}, \bibinfo
  {author} {\bibfnamefont {F.}~\bibnamefont {Copie}}, \bibinfo {author}
  {\bibfnamefont {G.}~\bibnamefont {Ducrozet}}, \bibinfo {author}
  {\bibfnamefont {A.}~\bibnamefont {Gelash}}, \bibinfo {author} {\bibfnamefont
  {G.}~\bibnamefont {Prabhudesai}}, \bibinfo {author} {\bibfnamefont
  {G.}~\bibnamefont {Michel}}, \bibinfo {author} {\bibfnamefont
  {A.}~\bibnamefont {Cazaubiel}}, \bibinfo {author} {\bibfnamefont
  {E.}~\bibnamefont {Falcon}}, \bibinfo {author} {\bibfnamefont
  {G.}~\bibnamefont {El}},\ and\ \bibinfo {author} {\bibfnamefont
  {S.}~\bibnamefont {Randoux}},\ }\bibfield  {title} {\enquote {\bibinfo
  {title} {Nonlinear spectral synthesis of soliton gas in deep-water surface
  gravity waves},}\ }\href@noop {} {\bibfield  {journal} {\bibinfo  {journal}
  {Phys. Rev. Lett.}\ }\textbf {\bibinfo {volume} {125}},\ \bibinfo {pages}
  {264101} (\bibinfo {year} {2020})}\BibitemShut {NoStop}%
\bibitem [{\citenamefont {Slunyaev}(2021)}]{Slunyaev2021}%
  \BibitemOpen
  \bibfield  {author} {\bibinfo {author} {\bibfnamefont {A.}~\bibnamefont
  {Slunyaev}},\ }\bibfield  {title} {\enquote {\bibinfo {title} {Persistence of
  hydrodynamic envelope solitons: detection and rogue wave occurrence},}\
  }\href@noop {} {\bibfield  {journal} {\bibinfo  {journal} {Phys. Fluids}\
  }\textbf {\bibinfo {volume} {33}},\ \bibinfo {pages} {036606} (\bibinfo
  {year} {2021})}\BibitemShut {NoStop}%
\bibitem [{\citenamefont {Slunyaev}, \citenamefont {Pelinovsky},\ and\
  \citenamefont {Pelinovsky}(2023)}]{Slunyaevetal2023}%
  \BibitemOpen
  \bibfield  {author} {\bibinfo {author} {\bibfnamefont {A.}~\bibnamefont
  {Slunyaev}}, \bibinfo {author} {\bibfnamefont {D.}~\bibnamefont
  {Pelinovsky}},\ and\ \bibinfo {author} {\bibfnamefont {E.}~\bibnamefont
  {Pelinovsky}},\ }\bibfield  {title} {\enquote {\bibinfo {title} {Rogue waves
  in the sea: observations, physics, and mathematics},}\ }\href@noop {}
  {\bibfield  {journal} {\bibinfo  {journal} {Physics--Uspekhi}\ }\textbf
  {\bibinfo {volume} {66}},\ \bibinfo {pages} {148--172} (\bibinfo {year}
  {2023})}\BibitemShut {NoStop}%
\bibitem [{\citenamefont {Zakharov}(1968)}]{Zakharov1968}%
  \BibitemOpen
  \bibfield  {author} {\bibinfo {author} {\bibfnamefont {V.}~\bibnamefont
  {Zakharov}},\ }\bibfield  {title} {\enquote {\bibinfo {title} {Stability of
  periodic waves of finite amplitude on a surface of deep fluid},}\ }\href@noop
  {} {\bibfield  {journal} {\bibinfo  {journal} {J. Appl. Mech. Tech. Phys.}\
  }\textbf {\bibinfo {volume} {2}},\ \bibinfo {pages} {190--194} (\bibinfo
  {year} {1968})}\BibitemShut {NoStop}%
\bibitem [{\citenamefont {West}\ \emph {et~al.}(1987)\citenamefont {West},
  \citenamefont {Brueckner}, \citenamefont {Janda}, \citenamefont {Milder},\
  and\ \citenamefont {Milton}}]{Westetal1987}%
  \BibitemOpen
  \bibfield  {author} {\bibinfo {author} {\bibfnamefont {B.}~\bibnamefont
  {West}}, \bibinfo {author} {\bibfnamefont {K.}~\bibnamefont {Brueckner}},
  \bibinfo {author} {\bibfnamefont {R.}~\bibnamefont {Janda}}, \bibinfo
  {author} {\bibfnamefont {D.}~\bibnamefont {Milder}},\ and\ \bibinfo {author}
  {\bibfnamefont {R.}~\bibnamefont {Milton}},\ }\bibfield  {title} {\enquote
  {\bibinfo {title} {A new numerical method for surface hydrodynamics},}\
  }\href@noop {} {\bibfield  {journal} {\bibinfo  {journal} {J. Geophys. Res.}\
  }\textbf {\bibinfo {volume} {92}},\ \bibinfo {pages} {11803--11824} (\bibinfo
  {year} {1987})}\BibitemShut {NoStop}%
\bibitem [{\citenamefont {Onorato}, \citenamefont {Osborne},\ and\
  \citenamefont {Serio}(2007)}]{Onoratoetal2007}%
  \BibitemOpen
  \bibfield  {author} {\bibinfo {author} {\bibfnamefont {M.}~\bibnamefont
  {Onorato}}, \bibinfo {author} {\bibfnamefont {A.}~\bibnamefont {Osborne}},\
  and\ \bibinfo {author} {\bibfnamefont {M.}~\bibnamefont {Serio}},\ }\bibfield
   {title} {\enquote {\bibinfo {title} {On the relation between two numerical
  methods for the computation of random surface gravity waves},}\ }\href@noop
  {} {\bibfield  {journal} {\bibinfo  {journal} {Eur. J. Mech. B/Fluids}\
  }\textbf {\bibinfo {volume} {26}},\ \bibinfo {pages} {43--48} (\bibinfo
  {year} {2007})}\BibitemShut {NoStop}%
\bibitem [{\citenamefont {Lavrova}(1983)}]{Lavrova1983}%
  \BibitemOpen
  \bibfield  {author} {\bibinfo {author} {\bibfnamefont {O.}~\bibnamefont
  {Lavrova}},\ }\bibfield  {title} {\enquote {\bibinfo {title} {On the lateral
  instability of waves on the surface of a fluid of finite depth},}\
  }\href@noop {} {\bibfield  {journal} {\bibinfo  {journal} {Izv. Atmos. Ocean.
  Phys.}\ }\textbf {\bibinfo {volume} {19}},\ \bibinfo {pages} {807--810}
  (\bibinfo {year} {1983})}\BibitemShut {NoStop}%
\bibitem [{\citenamefont {Slunyaev}(2009)}]{Slunyaev2009}%
  \BibitemOpen
  \bibfield  {author} {\bibinfo {author} {\bibfnamefont {A.}~\bibnamefont
  {Slunyaev}},\ }\bibfield  {title} {\enquote {\bibinfo {title} {Numerical
  simulation of limiting envelope solitons of gravity waves on deep water},}\
  }\href@noop {} {\bibfield  {journal} {\bibinfo  {journal} {JETP}\ }\textbf
  {\bibinfo {volume} {109}},\ \bibinfo {pages} {676--686} (\bibinfo {year}
  {2009})}\BibitemShut {NoStop}%
\bibitem [{\citenamefont {Peregrine}(1983)}]{Peregrine1983}%
  \BibitemOpen
  \bibfield  {author} {\bibinfo {author} {\bibfnamefont {D.}~\bibnamefont
  {Peregrine}},\ }\bibfield  {title} {\enquote {\bibinfo {title} {Water waves,
  nonlinear schrodinger equations and their solutions},}\ }\href@noop {}
  {\bibfield  {journal} {\bibinfo  {journal} {J. Austral. Math. Soc. Ser. B.}\
  }\textbf {\bibinfo {volume} {25}},\ \bibinfo {pages} {16--43} (\bibinfo
  {year} {1983})}\BibitemShut {NoStop}%
\bibitem [{\citenamefont {Akhmediev}\ and\ \citenamefont
  {Ankiewicz}(1993)}]{AkhmedievetalAnkiewicz1993}%
  \BibitemOpen
  \bibfield  {author} {\bibinfo {author} {\bibfnamefont {N.}~\bibnamefont
  {Akhmediev}}\ and\ \bibinfo {author} {\bibfnamefont {A.}~\bibnamefont
  {Ankiewicz}},\ }\bibfield  {title} {\enquote {\bibinfo {title} {Spatial
  soliton x-junctions and couplers},}\ }\href@noop {} {\bibfield  {journal}
  {\bibinfo  {journal} {Optics Communications}\ }\textbf {\bibinfo {volume}
  {100}},\ \bibinfo {pages} {186--192} (\bibinfo {year} {1993})}\BibitemShut
  {NoStop}%
\bibitem [{\citenamefont {Chabchoub}\ \emph {et~al.}(2021)\citenamefont
  {Chabchoub}, \citenamefont {Slunyaev}, \citenamefont {Hoffmann},
  \citenamefont {Dias}, \citenamefont {Kibler}, \citenamefont {Genty},
  \citenamefont {Dudley},\ and\ \citenamefont {Akhmediev}}]{Chabchoubetal2021}%
  \BibitemOpen
  \bibfield  {author} {\bibinfo {author} {\bibfnamefont {A.}~\bibnamefont
  {Chabchoub}}, \bibinfo {author} {\bibfnamefont {A.}~\bibnamefont {Slunyaev}},
  \bibinfo {author} {\bibfnamefont {N.}~\bibnamefont {Hoffmann}}, \bibinfo
  {author} {\bibfnamefont {F.}~\bibnamefont {Dias}}, \bibinfo {author}
  {\bibfnamefont {B.}~\bibnamefont {Kibler}}, \bibinfo {author} {\bibfnamefont
  {G.}~\bibnamefont {Genty}}, \bibinfo {author} {\bibfnamefont
  {J.}~\bibnamefont {Dudley}},\ and\ \bibinfo {author} {\bibfnamefont
  {N.}~\bibnamefont {Akhmediev}},\ }\bibfield  {title} {\enquote {\bibinfo
  {title} {The peregrine breather on the zero-background limit as the
  two-soliton degenerate solution: An experimental study},}\ }\href@noop {}
  {\bibfield  {journal} {\bibinfo  {journal} {Frontiers in Physics}\ }\textbf
  {\bibinfo {volume} {9}},\ \bibinfo {pages} {633549} (\bibinfo {year}
  {2021})}\BibitemShut {NoStop}%
\bibitem [{\citenamefont {Zakharov}\ and\ \citenamefont
  {Rubenchik}(1974)}]{ZakharovRubenchik1974}%
  \BibitemOpen
  \bibfield  {author} {\bibinfo {author} {\bibfnamefont {V.}~\bibnamefont
  {Zakharov}}\ and\ \bibinfo {author} {\bibfnamefont {A.}~\bibnamefont
  {Rubenchik}},\ }\bibfield  {title} {\enquote {\bibinfo {title} {Instability
  of waveguides and solitons in nonlinear media},}\ }\href@noop {} {\bibfield
  {journal} {\bibinfo  {journal} {Sov. Phys. JETP}\ }\textbf {\bibinfo {volume}
  {38}},\ \bibinfo {pages} {494--501} (\bibinfo {year} {1974})}\BibitemShut
  {NoStop}%
\bibitem [{\citenamefont {Slunyaev}\ and\ \citenamefont
  {Shrira}(2013)}]{SlunyaevShrira2013}%
  \BibitemOpen
  \bibfield  {author} {\bibinfo {author} {\bibfnamefont {A.}~\bibnamefont
  {Slunyaev}}\ and\ \bibinfo {author} {\bibfnamefont {V.}~\bibnamefont
  {Shrira}},\ }\bibfield  {title} {\enquote {\bibinfo {title} {On the highest
  non-breaking wave in a group: fully nonlinear water wave breathers vs weakly
  nonlinear theory},}\ }\href@noop {} {\bibfield  {journal} {\bibinfo
  {journal} {J. Fluid Mech.}\ }\textbf {\bibinfo {volume} {735}},\ \bibinfo
  {pages} {203--248} (\bibinfo {year} {2013})}\BibitemShut {NoStop}%
\bibitem [{\citenamefont {Slunyaev}\ \emph {et~al.}(2013)\citenamefont
  {Slunyaev}, \citenamefont {Pelinovsky}, \citenamefont {Sergeeva},
  \citenamefont {Chabchoub}, \citenamefont {Hoffmann}, \citenamefont
  {Onorato},\ and\ \citenamefont {Akhmediev}}]{Slunyaevetal2013PRE}%
  \BibitemOpen
  \bibfield  {author} {\bibinfo {author} {\bibfnamefont {A.}~\bibnamefont
  {Slunyaev}}, \bibinfo {author} {\bibfnamefont {E.}~\bibnamefont
  {Pelinovsky}}, \bibinfo {author} {\bibfnamefont {A.}~\bibnamefont
  {Sergeeva}}, \bibinfo {author} {\bibfnamefont {A.}~\bibnamefont {Chabchoub}},
  \bibinfo {author} {\bibfnamefont {N.}~\bibnamefont {Hoffmann}}, \bibinfo
  {author} {\bibfnamefont {M.}~\bibnamefont {Onorato}},\ and\ \bibinfo {author}
  {\bibfnamefont {N.}~\bibnamefont {Akhmediev}},\ }\bibfield  {title} {\enquote
  {\bibinfo {title} {Super rogue waves in simulations based on weakly nonlinear
  and fully nonlinear hydrodynamic equations},}\ }\href@noop {} {\bibfield
  {journal} {\bibinfo  {journal} {Phys. Rev. E}\ }\textbf {\bibinfo {volume}
  {88}},\ \bibinfo {pages} {012909} (\bibinfo {year} {2013})}\BibitemShut
  {NoStop}%
\bibitem [{\citenamefont {Slunyaev}, \citenamefont {Klein},\ and\ \citenamefont
  {Clauss}(2017)}]{Slunyaevetal2017}%
  \BibitemOpen
  \bibfield  {author} {\bibinfo {author} {\bibfnamefont {A.}~\bibnamefont
  {Slunyaev}}, \bibinfo {author} {\bibfnamefont {M.}~\bibnamefont {Klein}},\
  and\ \bibinfo {author} {\bibfnamefont {G.}~\bibnamefont {Clauss}},\
  }\bibfield  {title} {\enquote {\bibinfo {title} {Laboratory and numerical
  study of intense envelope solitons of water waves: generation, reflection
  from a wall and collisions},}\ }\href@noop {} {\bibfield  {journal} {\bibinfo
   {journal} {Phys. Fluids}\ }\textbf {\bibinfo {volume} {29}},\ \bibinfo
  {pages} {047103} (\bibinfo {year} {2017})}\BibitemShut {NoStop}%
\bibitem [{\citenamefont {He}\ \emph {et~al.}(2022)\citenamefont {He},
  \citenamefont {Slunyaev}, \citenamefont {Mori},\ and\ \citenamefont
  {Chabchoub}}]{Heetal2022}%
  \BibitemOpen
  \bibfield  {author} {\bibinfo {author} {\bibfnamefont {Y.}~\bibnamefont
  {He}}, \bibinfo {author} {\bibfnamefont {A.}~\bibnamefont {Slunyaev}},
  \bibinfo {author} {\bibfnamefont {N.}~\bibnamefont {Mori}},\ and\ \bibinfo
  {author} {\bibfnamefont {A.}~\bibnamefont {Chabchoub}},\ }\bibfield  {title}
  {\enquote {\bibinfo {title} {Experimental evidence of nonlinear focusing in
  standing water waves},}\ }\href@noop {} {\bibfield  {journal} {\bibinfo
  {journal} {Phys. Rev. Lett.}\ }\textbf {\bibinfo {volume} {129}},\ \bibinfo
  {pages} {144502} (\bibinfo {year} {2022})}\BibitemShut {NoStop}%
\end{thebibliography}%

\end{document}